\documentclass[a4paper, 11pt, oneside]{article} % A4 paper size, default 11pt font size and oneside for equal margins
\usepackage{epsfig}
\usepackage{authblk}
\usepackage[T1]{fontenc}
\usepackage{lmodern}
\usepackage{kpfonts}
\usepackage[hyphens]{url}
\usepackage{graphicx}%
\usepackage{multirow}%
\usepackage{amsmath,amssymb,amsfonts}%
\usepackage{amsthm}%
\usepackage{mathrsfs}%
\usepackage[title]{appendix}%
\usepackage{xcolor}%
\usepackage{textcomp}%
\usepackage{manyfoot}%
\usepackage{booktabs}%
\usepackage{algorithm}%
\usepackage{algorithmicx}%
\usepackage{algpseudocode}%
\usepackage{listings}%
\usepackage{comment}%
\usepackage[version=4]{mhchem}%
\usepackage{chemfig}%
\usepackage{subfigure}%
\usepackage{anyfontsize}%
\usepackage{adjustbox}%

\begin{document} 

%%%%封面内容编辑%%%%
\begin{titlepage} % Suppresses headers and footers on the title page

	\centering % Centre everything on the title page
	
	\scshape % Use small caps for all text on the title page
	
	\vspace*{\baselineskip} % White space at the top of the page
	
	%------------------------------------------------
	%	Title
	%------------------------------------------------
	
	\rule{\textwidth}{1.6pt}\vspace*{-\baselineskip}\vspace*{2pt} % Thick horizontal rule
	\rule{\textwidth}{0.4pt} % Thin horizontal rule
	
	\vspace{0.75\baselineskip} % Whitespace above the title
	
	{\LARGE On Meta-Evaluation} % Title
	
	\vspace{0.75\baselineskip} % Whitespace below the title
	
	\rule{\textwidth}{0.4pt}\vspace*{-\baselineskip}\vspace{3.2pt} % Thin horizontal rule
	\rule{\textwidth}{1.6pt} % Thick horizontal rule
	
	\vspace{2\baselineskip} % Whitespace after the title block
	
	%------------------------------------------------
	%	Subtitle
	%------------------------------------------------
	
	%Subtitle here % Subtitle or further description
	
	\vspace*{3\baselineskip} % Whitespace under the subtitle
	
	%------------------------------------------------
	%	Editor(s)
	%------------------------------------------------
	
	Edited By
	
	\vspace{0.5\baselineskip} % Whitespace before the editors
	
	{\scshape\Large Hongxiao Li\\ Chenxi Wang\\ Fanda Fan\\ Zihan Wang\\ Wanling Gao\\ Lei Wang\\ Jianfeng Zhan\\ }

	\vspace{0.5\baselineskip} % Whitespace below the editor list

	\vfill % Whitespace between editor names and publisher logo
	
	%------------------------------------------------
	%	Publisher
	%------------------------------------------------
	
	%\plogo % Publisher logo
	%\def\BUlogo{\epsfig{file=ICT.pdf,height=3cm}}
	%\includegraphics[scale=0.135]{ICT.pdf}
	\epsfig{file=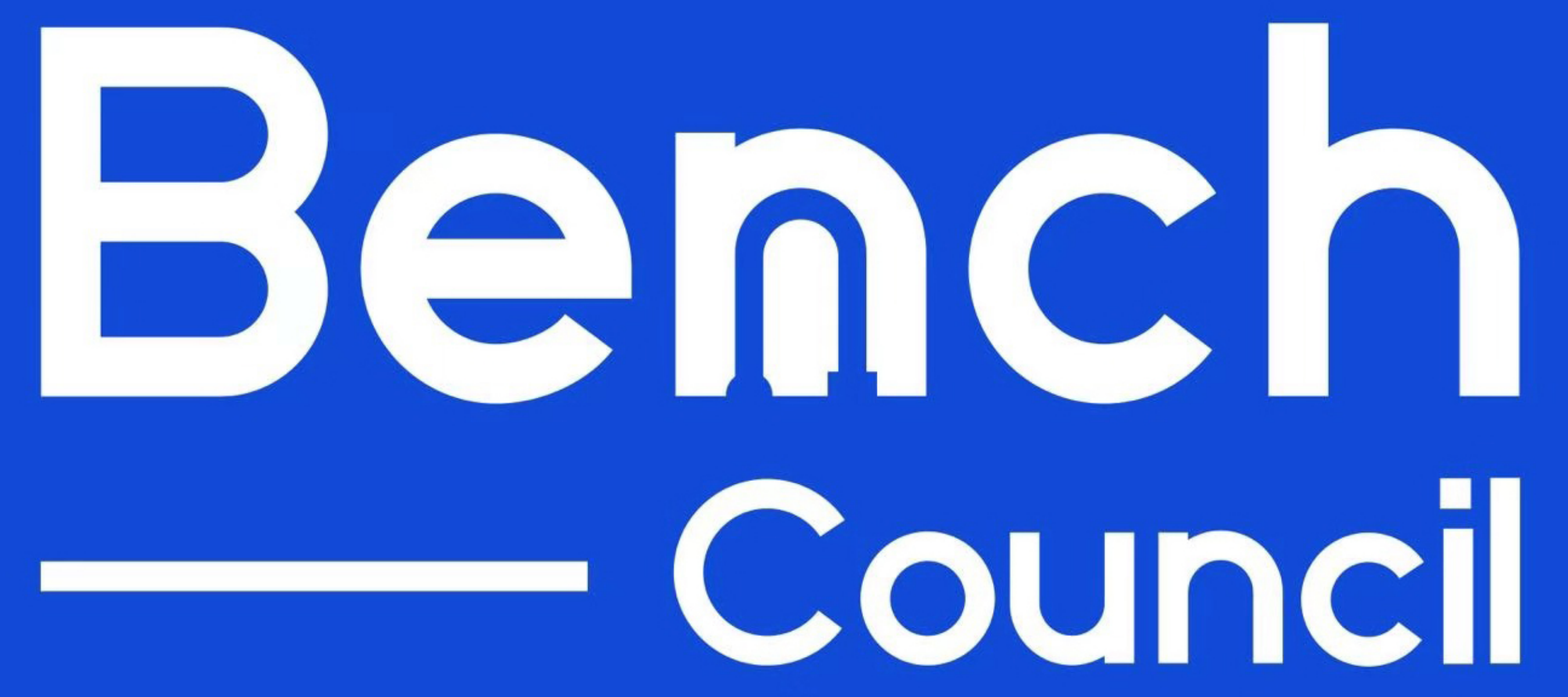,height=2cm}
	\\BenchCouncil: International Open Benchmark Council\\Chinese Academy of Sciences\\Beijing, China\\https://www.benchcouncil.org % Editor affiliation
	\vspace{5\baselineskip} % Whitespace under the publisher logo

	Technical Report No. BenchCouncil-OnMetaEvaluation-2025 % Publication year
	
	{\large Nov 27, 2025} % Publisher

\end{titlepage}

%----------------------------------------------------------------------------------------

%%%title here%%%
\title{On Meta-Evaluation}

\author[1,3]{Hongxiao Li}
\author[1,3]{Chenxi Wang}
\author[1,2,3]{Fanda Fan}
\author[1]{Zihan Wang}
\author[1,2,3]{Wanling Gao}
\author[1,2,3]{Lei Wang}
\author[1,2,3]{Jianfeng Zhan\thanks{Jianfeng Zhan is the corresponding author.}}

\affil[1]{Institute of Computing Technology, Chinese Academy of Sciences\\ \{wangchenxi21s, fanfanda, gaowanling, wanglei\_2011, zhanjianfeng\}@ict.ac.cn, lihongxiao19@mails.ucas.ac.cn, zihanWang@ahnu.edu.cn}
\affil[2]{BenchCouncil (International Open Benchmark Council)}
\affil[3]{University of Chinese Academy of Sciences}

\date{Nov 27, 2025}
\maketitle

\begin{abstract}

Evaluation is the foundation of empirical science, yet the evaluation of evaluation itself -- so-called meta-evaluation -- remains strikingly underdeveloped. While methods such as observational studies, design of experiments (DoE), and randomized controlled trials (RCTs) have shaped modern scientific practice, there has been little systematic inquiry into their comparative validity and utility across domains. Here we introduce a formal framework for meta-evaluation by defining the evaluation space, its structured representation, and a benchmark we call AxiaBench. AxiaBench enables the first large-scale, quantitative comparison of ten widely used evaluation methods across eight representative application domains. Our analysis reveals a fundamental limitation: no existing method simultaneously achieves accuracy and efficiency across diverse scenarios, with DoE and observational designs in particular showing significant deviations from real-world ground truth. We further evaluate a unified method of entire-space stratified sampling from previous evaluatology research, and the results report that it consistently outperforms prior approaches across all tested domains. These results establish meta-evaluation as a scientific object in its own right and provide both a conceptual foundation and a pragmatic tool set for advancing trustworthy evaluation in computational and experimental research.

\end{abstract}

\section{Introduction}\label{sec:introduction}

%Evaluation holds significant importance across various fields of study. In the science history, researchers propose various evaluation methods across different sub-domains including physics, chemistry, informatics, and so on. There are different evaluation methods along the scientific research history, including but not limited to design of experiments (DoE)~\cite{jain1990art,douglas2001design} in agricultural, randomized controlled trials (RCTs)~\cite{chalmers1981method} in medical, causal inference (CI)~\cite{imbens2015causal} in environment science, and quasi-experiment~\cite{thyer2012quasi} in social science. Besides, most research adopts an observation~\cite{zalta2003stanford} at least. The recent Nobel Prizes have all been awarded to researchers in recognition of their groundbreaking work in the field of computer science.

Evaluation is the scaffolding on which empirical science is built. From physics to medicine, from economics to computer science, scientific progress has been enabled not only by discoveries themselves but also by the methods through which evidence is gathered, tested, and judged. Across history, diverse approaches have emerged to serve this role: the design of experiments (DoE)~\cite{jain1990art,douglas2001design} in agricultural science, randomized controlled trials (RCTs)~\cite{chalmers1981method} in medicine, causal inference frameworks~\cite{imbens2015causal} in social and environmental studies, and quasi-experiments~\cite{thyer2012quasi} in economics and the behavioral sciences. At the very least, observational designs~\cite{zalta2003stanford} permeate nearly every empirical discipline. Recent Nobel Prizes in the sciences underscore the centrality of evaluation in shaping rigorous and trustworthy knowledge~\cite{jumper2021highly}.

%Unfortunately, the formal definition and pragmatical paradigm have not been unified in consensus, let alone the evaluation of evaluation -- meta evaluation. The concept of meta evaluation is firstly proposed in the work of Scriven~\cite{scriven1991evaluation}. However, this work provides only basic concepts and rules instead of concrete design and practical utility for meta evaluation. Besides, there are existing studies on some mature evaluation methods, including observation, design of experiments (DoE), randomized controlled trials (RCTs), reporting possible effectiveness and limitations of different methods, not including the comparison of their performances with several definite indexes in any certain concrete scenarios. For example, there is a CPU performance evaluation work~\cite{wang2024achieving} that involves similar concerns.

Yet despite its foundational role, evaluation itself remains curiously undertheorized. While methodological traditions have crystallized within individual domains, the formal study of meta-evaluation -- the evaluation of evaluation -- has scarcely advanced beyond conceptual sketches. Scriven’s early articulation~\cite{scriven1991evaluation} provided only definitions and guiding principles, without offering systematic design, operational frameworks, or practical toolsets. Existing studies typically highlight the strengths and weaknesses of individual methods -- such as DoE~\cite{jain1990art,douglas2001design}, RCTs~\cite{chalmers1981method}, or observational studies~\cite{zalta2003stanford} -- but rarely provide quantitative comparisons across contexts, nor do they define a general space in which evaluation methods can be analyzed on common ground. As a result, the practice of evaluation remains fragmented, and the choice of method is often dictated by disciplinary convention rather than empirical validity or efficiency.
Recent attempts to formalize evaluation theory, such as Zhan et al.'s ``Evaluatology''~\cite{zhan2024evaluatology} and a CPU performance evaluation work~\cite{wang2024achieving} represent an important step forward, introducing notions such as evaluation conditions and minimal evaluation systems. However, even this work stops short of addressing the core challenge: how can we systematically measure, compare, and generalize the effectiveness of evaluation methods across diverse scientific and engineering domains?

In this work, we advance that agenda by introducing a formal framework for meta-evaluation. We define the evaluation condition (EC) space, specify its structured representation, and instantiate it through a benchmark we call AxiaBench. For every task, the EC space includes several multiple-value variables that stand for the concrete configurations of the evaluation and have essential influence on the evaluation outcomes. This framework enables, for the first time, a systematic and quantitative comparison of ten representative evaluation methods
%—including observation, multiple DoE variants~\cite{jain1990art}, classical RCTs~\cite{chalmers1981method}, do-calculus~\cite{imbens2015causal,powell2018book} and Structural Causal Models (SCM)\cite{imbens2015causal,powell2018book} from causal inference, as well as non-randomized controlled trials (NRCTs)\cite{thyer2012quasi} and intervention-staggering approaches~\cite{thyer2012quasi} from quasi-experimental designs. 
%These methods are evaluated 
across eight canonical tasks spanning the natural, computational, and social sciences.
%: celestial motion prediction~\cite{lichtenberg2013regular,cambel1993applied}, rainfall prediction~\cite{burrows1991objective}, population dynamics simulation~\cite{turing1990chemical,tilman1997spatial}, formation energy prediction~\cite{wicks1963thermodynamic,kohn1965self}, pseudo-random number verification~\cite{may1976simple,rukhin2001statistical}, revenue assessment of game strategies~\cite{axelrod1981evolution}, CPU performance evaluation~\cite{wang2024achieving}, and LLM-based problem solving~\cite{yang2023gpt,liu2024finemath}. 
To assess the usability of evaluation methods, we focus on two fundamental criteria: accuracy and cost. %Our analysis uncovers a critical limitation: no existing method achieves both accuracy and cost-efficiency across all scenarios, with DoE-based approaches and observational designs in particular exhibiting systematic deviations from real-world ground truth.

%The results provide three insights as follows. Firstly, cost-tolerant researchers usually achieve better accuracy on evaluations, because RCTs method~\cite{chalmers1981method}, causal method~\cite{imbens2015causal,powell2018book}, and our space-stratified sampling method outperforms all other evaluation methods at over 95\% accuracy, if regardless of cost. Especially, our space-stratified sampling method achieves the same accuracy but takes half less cost on most tasks. Secondly, for the tasks that can adopt the observation~\cite{zalta2003stanford}, the best performance of the observation is far less than other methods, sometimes even lower than 50\% accuracy and can lead to mistaken conclusions. More severely, there is accuracy descend on the observation, which means experiment repetition aggravates biases. Thirdly, the $2^kr$ DoE method~\cite{jain1990art} and some quasi-experiment method~\cite{thyer2012quasi} work well on half tasks (over 90\% accuracy) with a loose estimation requirement, but under-performs with a strict estimation requirement. Besides, DoE performs at 60\% or lower accuracy on several tasks, which means that it is not feasible for evaluation on such scenarios.

Our results yield three key insights. First, accuracy tends to increase when cost is not a constraint: classical RCTs~\cite{chalmers1981method}, and causal inference methods~\cite{imbens2015causal,powell2018book} all exceed 95\% accuracy, with the best method achieving comparable performance at roughly half the cost. Second, observational designs~\cite{zalta2003stanford} perform markedly worse, often falling below 50\% accuracy and, critically, degrading further with repeated trials, thereby amplifying rather than reducing bias. Third, certain DoE variants (e.g., $2^kr$\cite{jain1990art}) and quasi-experimental designs\cite{thyer2012quasi} achieve acceptable accuracy ($>$ 90\%) under relaxed estimation requirements, but their reliability collapses under stricter criteria, with accuracies dropping to 60\% or lower in several domains.

%This work is the first systematical analysis on practical meta evaluation. We argue that the conclusion still holds true in emerging fields, because we experiments achieved similar results on tasks from different sub-fields. We developed our methodology into a practical evaluation suite -- AxiaBench, aiming at assessing the advantages and disadvantages of the evaluation methods in different sub-fields.

Taken together, these findings constitute the first systematic, cross-domain analysis of practical meta-evaluation. Importantly, the conclusions generalize to emerging fields, as validated by consistent results across diverse scientific tasks. %Building on this foundation, we operationalize our methodology as AxiaBench, a unified evaluation suite that reveals the strengths and limitations of competing methods, thereby 
%The methodology and its instantiation benchmark called AxiaBench position meta-evaluation not merely as a methodological exercise, but as a cornerstone of scientific inquiry.
The methodology and its instantiation, AxiaBench, elevate meta-evaluation from a methodological exercise to a cornerstone of scientific inquiry, providing both conceptual grounding and practical utility.

\begin{figure}
	\centering
		\includegraphics[scale=.5]{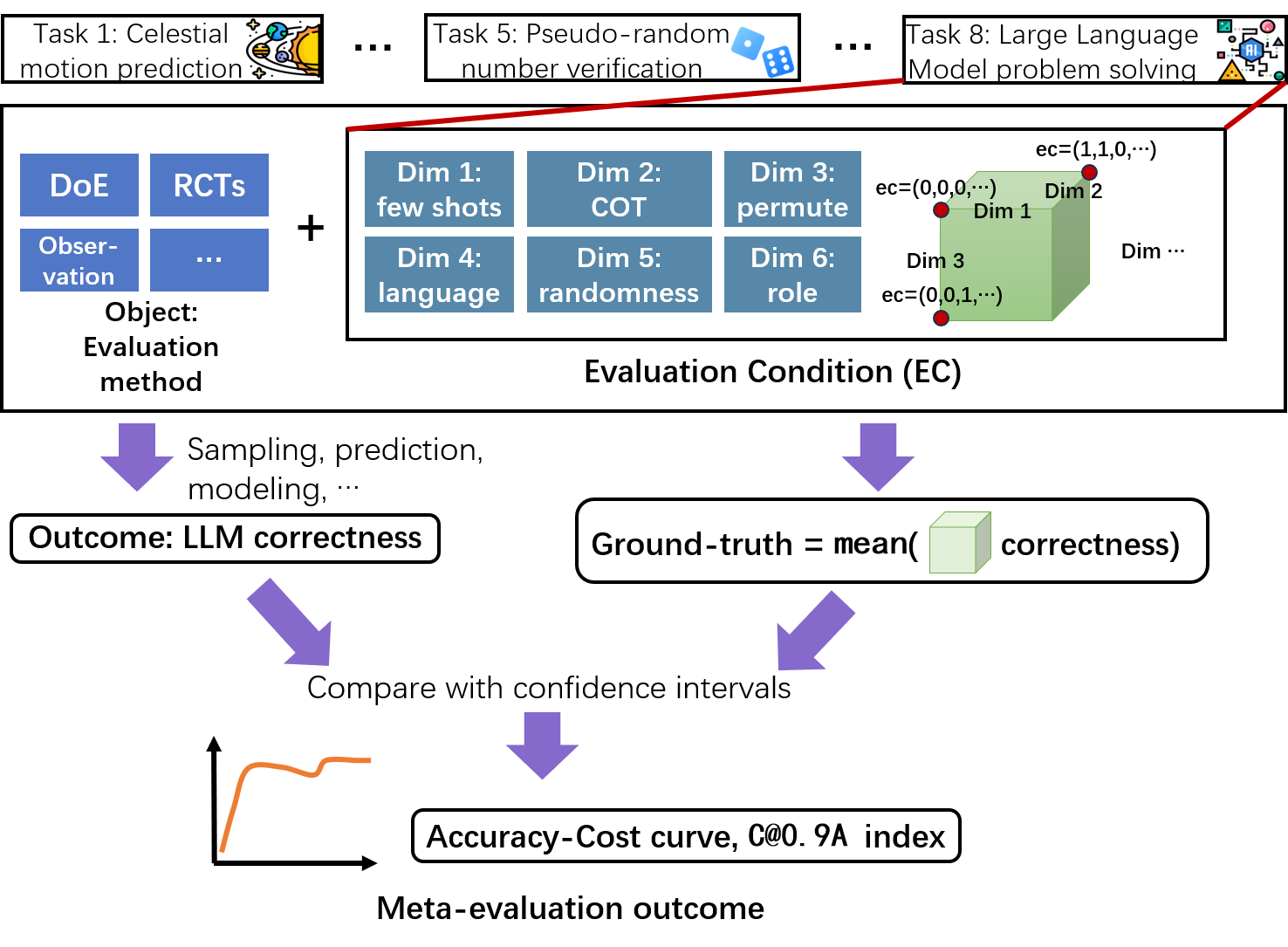}
	\caption{The complete framework of meta-evaluation involves the assessment of a specific evaluation methodology, shown using Large Language Model evaluation task as the example. In meta-evaluation, the EC space is that of the original EC along with the original evaluation object, while the meta-evaluation object is a certain evaluation method. The ground truth is established as the mean outcome across the entire spectrum (i.e. six dimensions in the example). Subsequently, the performance of the methodology is juxtaposed with the ground truth, reflecting both its precision and efficiency.}
	\label{fig:motivation}
\end{figure}

Inspired by evaluatology~\cite{zhan2024evaluatology}, we evaluate an entire-space stratified sampling method (also the evaluatology method for short) from previous evaluatology research~\cite{wang2024achieving}. Applied through AxiaBench, this approach consistently achieves the best balance of accuracy and cost across all tasks. Beyond its immediate utility, it gestures toward a broader vision: that evaluation itself can be theorized, systematized, and optimized as a universal science of evidence. By releasing all tools, datasets, and results openly, we aim to foster a community-wide effort to transform evaluation from a fragmented practice into a coherent framework that underwrites the reliability of knowledge across disciplines (Fig.~\ref{fig:motivation}).

\section{Results}\label{sec:results}

We conducted meta-evaluation experiments across eight distinct tasks: celestial motion prediction~\cite{lichtenberg2013regular,cambel1993applied}, rainfall prediction~\cite{burrows1991objective}, population dynamics simulation~\cite{turing1990chemical,tilman1997spatial}, formation energy prediction~\cite{wicks1963thermodynamic,kohn1965self}, pseudo-random number verification~\cite{may1976simple,rukhin2001statistical}, revenue evaluation of game strategies~\cite{axelrod1981evolution}, CPU performance evaluation~\cite{wang2024achieving}, and large language model (LLM) problem solving~\cite{yang2023gpt,liu2024finemath}.

The importance of evaluations in these fields cannot be overstated. In the realm of intelligent prediction, evaluations serve to validate essential physical laws and fine-tune model parameters for improved accuracy. They also aid in gaining a deeper insight into the mechanisms of intricate systems. In material science, establishing reliable evaluations can significantly expedite the discovery of novel materials. Evaluation in numerical and strategic simulation enables safeguarding the integrity of various applications and the dependability of experimental outcomes. Moreover, it helps in pinpointing robust strategies and serves as a foundation for making optimization decisions. In computer hardware and large language model domains, evaluations offer valuable benchmarks for stakeholders involved in development and procurement processes.

The evaluation methodologies assessed cover most state-of-the-art and state-of-the-practice methods across multiple domains: observation, $2^k$, $2^{k-p}$, $2^kr$, $l^kr$ methods of DoE~\cite{jain1990art} classical RCTs~\cite{chalmers1981method}, do-calculus~\cite{imbens2015causal,powell2018book} of causal inference, Structural Causal Model (SCM)~\cite{imbens2015causal,powell2018book}, non-randomized controlled trials (NRCTs)~\cite{thyer2012quasi} in quasi-experiment, intervention staggering method~\cite{thyer2012quasi} in quasi-experiment, and the evaluatology method~\cite{wang2024achieving} in recent CPU evaluation. The abbreviation of them are as follows, as appearance in late results figures and tables. Observation: \texttt{Obs}, $2^k$: \texttt{Doe\_2\textasciicircum k}, $2^kr$: \texttt{Doe\_2\textasciicircum kr}, $2^{k-p}$: \texttt{Doe\_2\textasciicircum(k-p)}, $l^k$: \texttt{Doe\_l\textasciicircum k}, RCTs: \texttt{Rct}, do-calculus: \texttt{Ci\_do}, SCM: \texttt{Ci\_scm}, NRCTs: \texttt{Qe\_nrct}, intervention staggering: \texttt{Qe\_stagger}, and evaluatology method: \texttt{Eva}.

\subsection{Celestial motion prediction}

This task is defined as evaluating the degree of chaos of different Lorenz system equations in the atmospheric convection model~\cite{emanuel1994atmospheric}, where different strange attractors are used to simulate real-world atmosphere chaotic systems, parameters and initial values of which representing various atmospheric conditions. We adopt the Lorentz attractor~\cite{lorenz2017deterministic} and R\"ossler attractor~\cite{rossler1983chaotic} for simulation and evaluation under several precisions and simulation steps, using the following two indexes for the degree of chaos: Lyapunov exponent~\cite{dingwell2006lyapunov}, and Kolmogorov-Smirnov (KS) statistics~\cite{fasano1987multidimensional}. This experiment aims at evaluating the indexes' accuracy estimation ability of different evaluation methods. The detailed illustration of this task is in Section~\ref{sec:methods}.

Our experimental results indicate that the $2^k$ DoE method achieves over 95\% accuracy with fewer than 20 evaluations for both the KS statistic and the Lyapunov exponent. In contrast, an $l^k$ DoE approach with $l=4$ shows no improvement in accuracy despite a 16-fold increase in cost, and performance declines to below 80\% accuracy even after 200 evaluations, with a maximum accuracy not exceeding 95\%. This suggests that the $l^k$ method is unsuitable for this application.

The observation method performed poorly, with accuracy below 40\% for the KS statistic at 30 evaluations and no improvement with increased cost. For the Lyapunov exponent, accuracy reached 90\% for the Lorenz attractor but only 44\% for the R\"ossler attractor after 100 repetitions.

Both the RCTs and evaluatology methods achieved the highest performance, converging to 98\% accuracy within 60 evaluations. The SCM method, when applied to the full variable space, performed equivalently to do-calculus. A limited evaluation of a partial variable space resulted in either 100\% or 0\% accuracy, depending on the configuration.

\begin{figure*}
\centering
    \subfigure[Task 1]{%
      \label{fig:task1}
      \centering
      \includegraphics[width=0.9\textwidth]{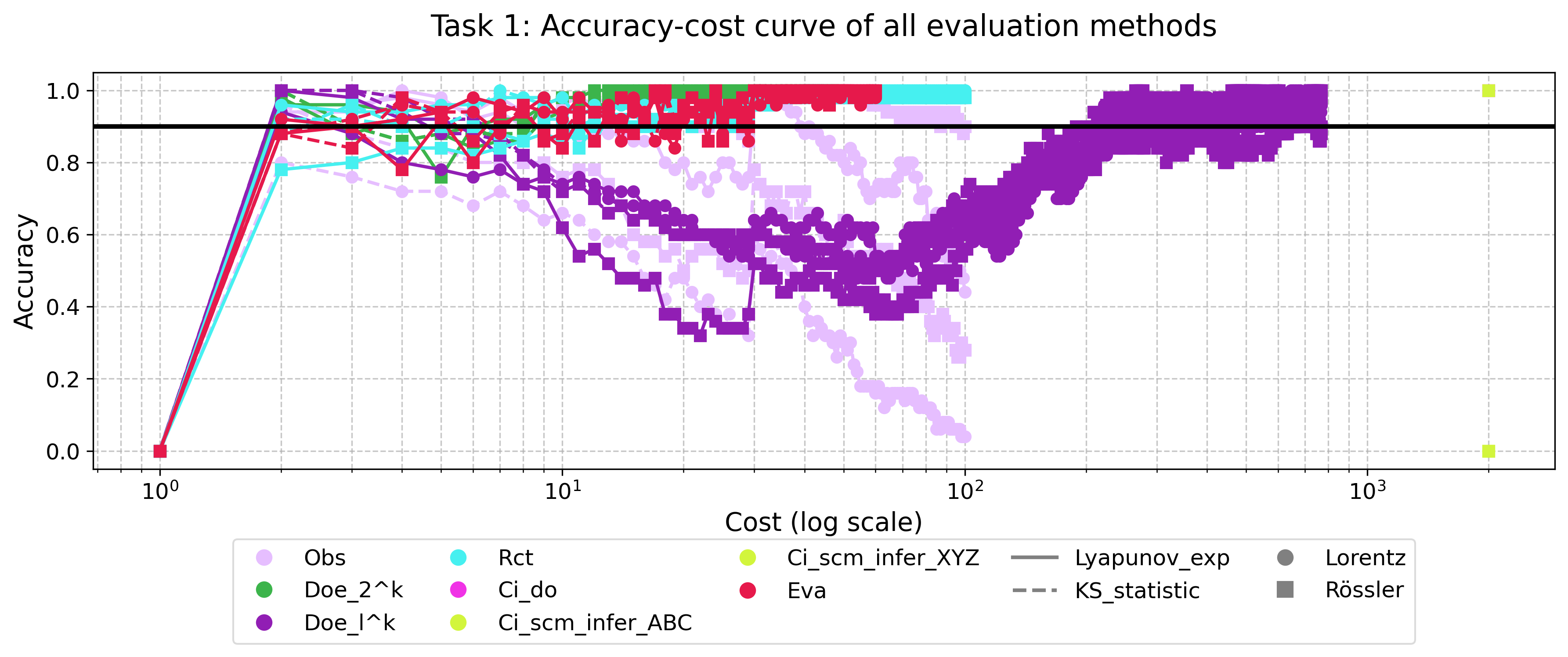}
    }
    \vspace{10pt}
    \subfigure[Task 2]{%
      \label{fig:task2}
      \centering
      \includegraphics[width=0.9\textwidth]{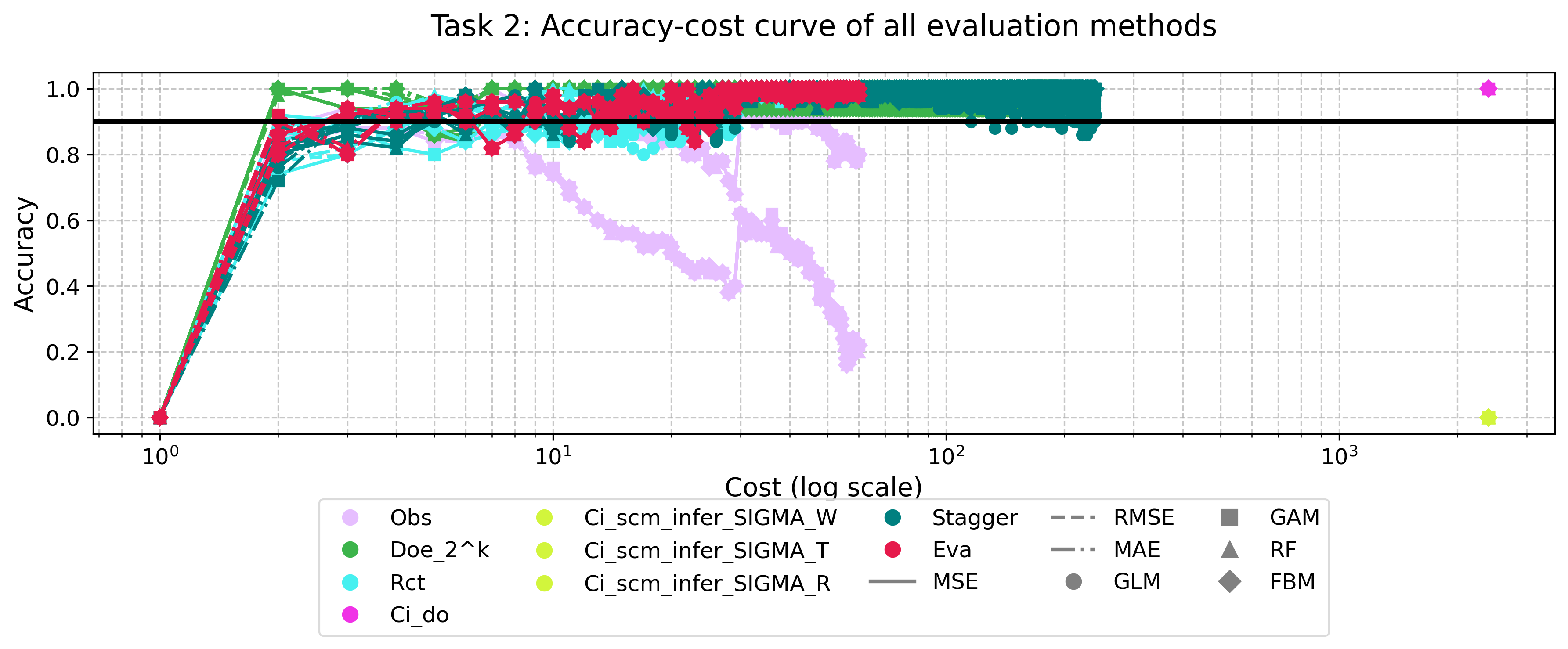}
    }
    \vspace{10pt}
    \subfigure[Task 3]{%
      \label{fig:task3}
      \centering
      \includegraphics[width=0.9\textwidth]{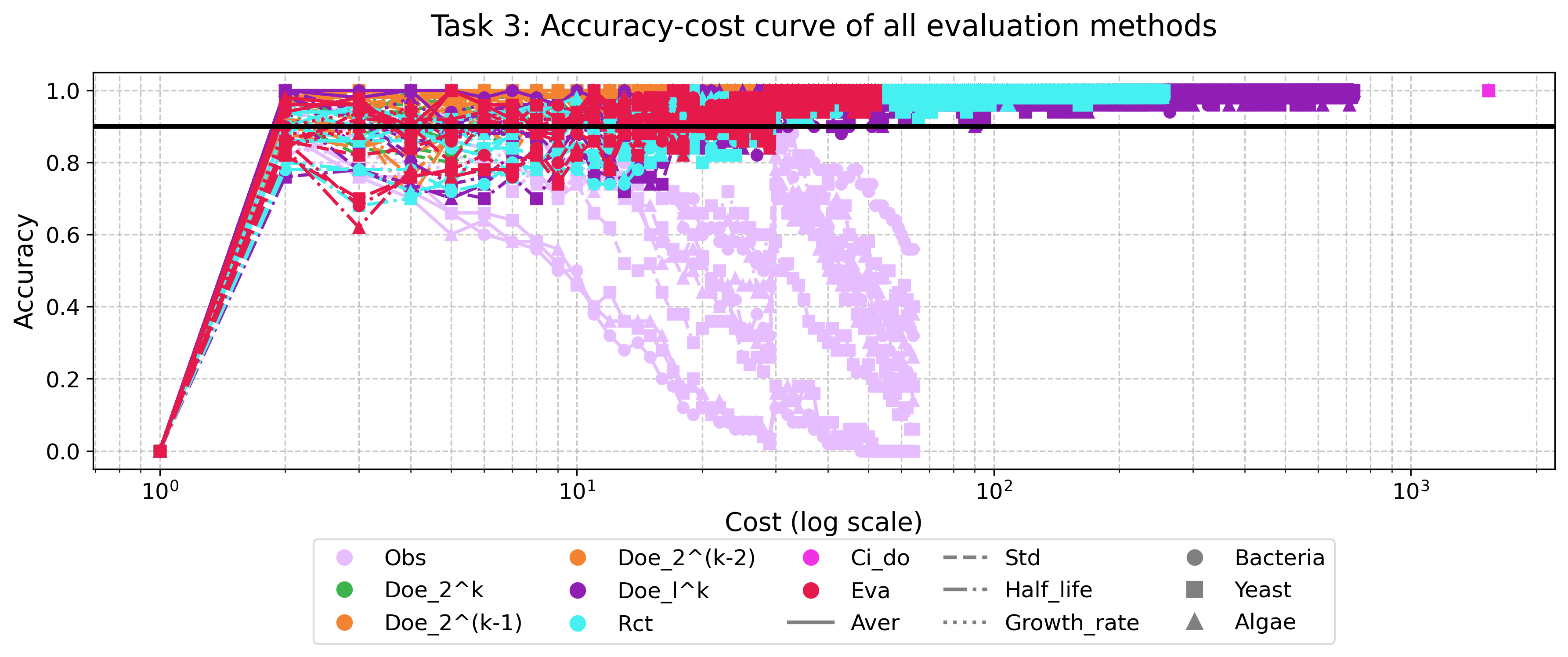}
    }
\caption{The accuracy–cost curves for various evaluation methods in Tasks 1, 2 and 3 are shown above. Colors represent different methods, shapes correspond to objects, and line styles denote evaluation metrics. The red lines indicate the Evaluatology method, which achieves the highest performance.}
\label{fig:task123}
\end{figure*}

In conclusion, RCTs and evaluatology are optimal and statistically indistinguishable for this task. The $2^k$ DoE method remains viable, whereas the $l^k$ method with $l=4$ is not recommended. Accuracy–cost curves for all methods are presented in Fig.~\ref{fig:task1}.

\subsection{Rainfall prediction}

This task is defined as evaluating the accuracy of the regression models in precipitation prediction problems, using a simplified \textit{ab initio} dataset~\cite{clausius1850ueber,emanuel1994atmospheric} simulating the real-world annual rainfall condition of a certain city. We evaluate four classical regression models~\cite{zuur2009glm}: GLM, GAM, RF, and FBM, under time series data computed based on several saturation specific humidity, vertical airflow intensity, temperature, airflow noise, temperature noise, observation noise values. The evaluation indexes taken are MSE, RMSE, and MAE. This experiment aims at evaluating the indexes' accuracy estimation ability of different evaluation methods. The detailed illustration of this task is in Section~\ref{sec:methods}.

%Our experiment results of this task is as follows. First, the $2^k$ DoE method achieves the complete 100\% accuracy at the cost of no more than 10 times of evaluations for both the MAE index and the RMSE index, and it achieves the 94\% accuracy for the MSE index. Among all eight tasks, DoE method at this task shows the best performance. Due to the EC space shape of this task, there is no feasible $l^k$ DoE methods. Second, the observation method's accuracies rapidly descend to 20\% while the times of repetition increase. For the MAE index and the RMSE index, the accuracies at cost of 100 times converge to 20\%, while the accuracy for the MSE index converges to 20\% at the cost of about 60 times of evaluation. This shows that observation totally failed at evaluation. Third, the results for the RCTs method and the evaluatology method achieve the accuracies that converge to 98\% accuracies within the cost of no more than 60 times of evaluation. The SCM method is equivalent to do-calculus, if the entire EC space is evaluated through. As a complement, an experiment of the SCM method of partial space is set. The results report that it all failed in evaluation, at the accuracy of 0\%. The intervention staggering method is inferior than the RCTs and evaluatology method, at the maximum 90\% accuracy. As a conclusion, using $2^k$ DoE method, the RCTs, or the evaluatology method should be considered as the optimal method and taking which one has no significant impact. Figure~\ref{fig:task2} shows the accuracy-cost curve of different evaluation methods in this task.

Our experimental results for this task are summarized as follows. First, the $2^k$ DoE approach achieved 100\% accuracy in both the MAE and RMSE indices with fewer than 10 evaluations, and attained 94\% accuracy for the MSE index. Among all eight tasks, the DoE method performed most effectively here. Owing to the structure of the EC space in this task, no feasible $l^k$ DoE configurations could be applied.

Second, the observation method exhibited a rapid decline in accuracy, converging to approximately 20\% as the number of repetitions increased. For the MAE and RMSE indices, accuracy stabilized near 20\% after 100 evaluations, whereas for the MSE index, it reached this level after approximately 60 evaluations, indicating a comprehensive failure.

Third, both the RCTs and evaluatology methods achieved accuracies converging to 98\% within 60 evaluations. The SCM method, equivalent to do-calculus when exhaustively evaluating the entire EC space, was also tested on a partial subspace for comparison. Results from this partial evaluation showed complete failure, with accuracies consistently at 0\%. The intervention staggering method underperformed relative to both RCTs and evaluatology, reaching a maximum accuracy of only 90\%.

In conclusion, the $2^k$ DoE, RCTs, and evaluatology methods are optimal choices for this task, with no significant performance differences observed among them. Accuracy–cost curves for each method are presented in Figure~\ref{fig:task2}.

\subsection{Population dynamics simulation}

This task is defined as evaluating population dynamic simulation based on cellular automata in a two-dimensional grid, where each cell can represent an individual of a species, where the state of the cell describes whether there is an individual at that location. The expansion and competition of a population can be simulated according to certain rules, such as growth, reproduction and death. We simulate the population size changes of bacteria~\cite{ben2000cooperative}, yeast~\cite{pfeiffer2001cooperation} and algae~\cite{huisman1999biodiversity}, that the simulation rules for each are slightly different. We adopt the following indexes used to evaluate the stability of the population in the cellular automaton model: average population, standard deviation of the population, half-life, and growth rate~\cite{murray2007mathematical}. We put forward evaluation under different initial distributions, simulation steps, environment capacity, and resource concentration degrees. This experiment aims at evaluating the indexes' accuracy estimation ability of different evaluation methods. The detailed illustration of this task is in Section~\ref{sec:methods}.

%Our experiment results of this task is as follows. First, the $2^k$ DoE method achieves over 90\% accuracy at the cost of 30 times of evaluations. This conclusion is the same for the predictions on the growth of bacteria, yeast, and algae. For the average population, standard deviation of the population, and growth rate indexes, the accuracy is up to 95\%. The $2^{k-p}$ method achieves the same accuracy at a lower cost, taking $p=1$ and $p=2$. However, the $l^k$ DoE method taking $l=4$ has no accuracy improvement but a 8-times cost. Therefore, it is not a feasible evaluation method for this task. Second, the observation method achieves unbearable low performance, while taking more evaluation cost by repetition results in even worse performance. For the average population index, the accuracy at cost of 30 times is close to 0\%, and there is no improvement when increasing evaluation cost. The accuracies for the growth rate and half-life indexes converge to about 20\% to 40\%. The accuracy for standard deviation index converges to lower than 20\%, where bacteria's is slightly better than algae's, and then yeast's. Third, the results for the RCTs method and the evaluatology method is generally the same as those of the $2^k$ DoE method. As a conclusion, using $2^k$, $2^{k-p}$, RCTs, or the evaluatology method should be considered as the optimal method and taking which one has no significant impact. Figure~\ref{fig:task3} shows the accuracy-cost curve of different evaluation methods in this task.

Our experimental findings for this task are summarized as follows. First, the $2^k$ full-factorial DoE method consistently achieved accuracy exceeding 90\% across all metrics -- bacteria, yeast, and algae growth -- within 30 evaluations. Specifically, for the average population, population standard deviation, and growth rate indices, accuracy reached up to 95\%. The $2^{k-p}$ factorial design attained comparable performance at reduced cost for $p=1$ and $p=2$. In contrast, an $l^k$ DoE configuration with $l=4$ showed no improvement in accuracy despite an approximately eightfold increase in evaluation cost, rendering it impractical for this application.

Second, the observation method exhibited consistently poor performance. Accuracy for the average population fell to nearly 0\% after 30 evaluations, with no improvement upon more repetition. For growth rate and half-life indices, accuracy plateaued between 20\% and 40\%, while the standard deviation metric converged below 20\%—with marginally better results for bacteria than algae or yeast.

Third, both RCTs and evaluatology methods delivered performance equivalent to the $2^k$ DoE approach, achieving high accuracy within a constrained evaluation cost concern.

In conclusion, the $2^k$, $2^{k-p}$, RCTs, and evaluatology methods all represent robust and efficient strategies for this task, with no statistically significant differences in outcome. Accuracy–cost curves for each method are presented in Figure~\ref{fig:task3}.

\subsection{Formation energy prediction}

This task is defined as formation energy prediction~\cite{peterson2021materials} evaluation using classical machine learning algorithms. This task include several different patterns of materials, described by general molecule modules and different radicals made of different chemical elements. We simulate the formation energies of materials by computation using \textit{ab initio} methods within a certain range of errors. The objects evaluated in this task include several state-of-the-art models. We adopt the following two indexes used to evaluate the accuracies of objects: MAPE, and SMAPE. This experiment aims at evaluating the indexes' accuracy estimation ability of different evaluation methods. The detailed illustration of this task is in Section~\ref{sec:methods}.

%Our experiment results of this task is as follows. The RCTs method and the evaluatology method achieve over 80\% accuracy and over 90\% accuracy at the cost of 60 times of evaluations for MAPE and SMAPE indexes respectively. Besides, there are no more improvements when taking more evaluation times cost. The results for the four AI models (i.e., PRB'14~\cite{meredig2014combinatorial}, ICDM'16~\cite{agrawal2016formation}, LR, and SVM) are no significant differences. The evaluatology method is slightly better compared to the RCTs method, due to less evaluation cost at the same accuracy performances. DoE methods are not suitable for this task, because the EC values are not completely independent, and there are some materials that cannot exist. The NRCTs method achieves similar but slightly lower performance to that of RCTs. In details, for the MAPE index, the accuracy is about 60\% at the cost of 30 evaluations, and it ascends to 90\% at the cost of 60. For the SMAPE index, they are about 75\% and 90\% at the cost of 30 and 60 respectively. As a conclusion, using RCTs, or the evaluatology method should be considered as a feasible practice and the evaluatology method is better at cost. The NRCTs method is also an inferior however feasible alternative approach for this task. Figure~\ref{fig:task4} shows the accuracy-cost curve of different evaluation methods in this task.

Our experimental results for this task are summarized as follows. Both the RCTs and the evaluatology method achieved high performance, exceeding 80\% accuracy for the MAPE index and 90\% for the SMAPE index, within 60 evaluations. No further improvements were observed with additional evaluations. Performance across the four AI models -- PRB'14~\cite{meredig2014combinatorial}, ICDM'16~\cite{agrawal2016formation}, logistic regression (LR), and support vector machine (SVM) -- showed no significant differences. The evaluatology method demonstrated a marginal advantage over RCTs in terms of evaluation efficiency, achieving comparable accuracy at a lower cost.

DoE methods were deemed unsuitable for this task due to dependencies among EC values and the presence of infeasible material combinations.

The non-randomized controlled trials (NRCTs) method exhibited slightly inferior but acceptable performance relative to RCTs. Specifically, for the MAPE index, NRCTs reached approximately 60\% accuracy at 30 evaluations and improved to 90\% at 60 evaluations. Corresponding values for the SMAPE index were approximately 75\% and 90\%, respectively.

\begin{figure*}
\centering
    \subfigure[Task 4]{%
      \label{fig:task4}
      \centering
      \includegraphics[width=0.9\textwidth]{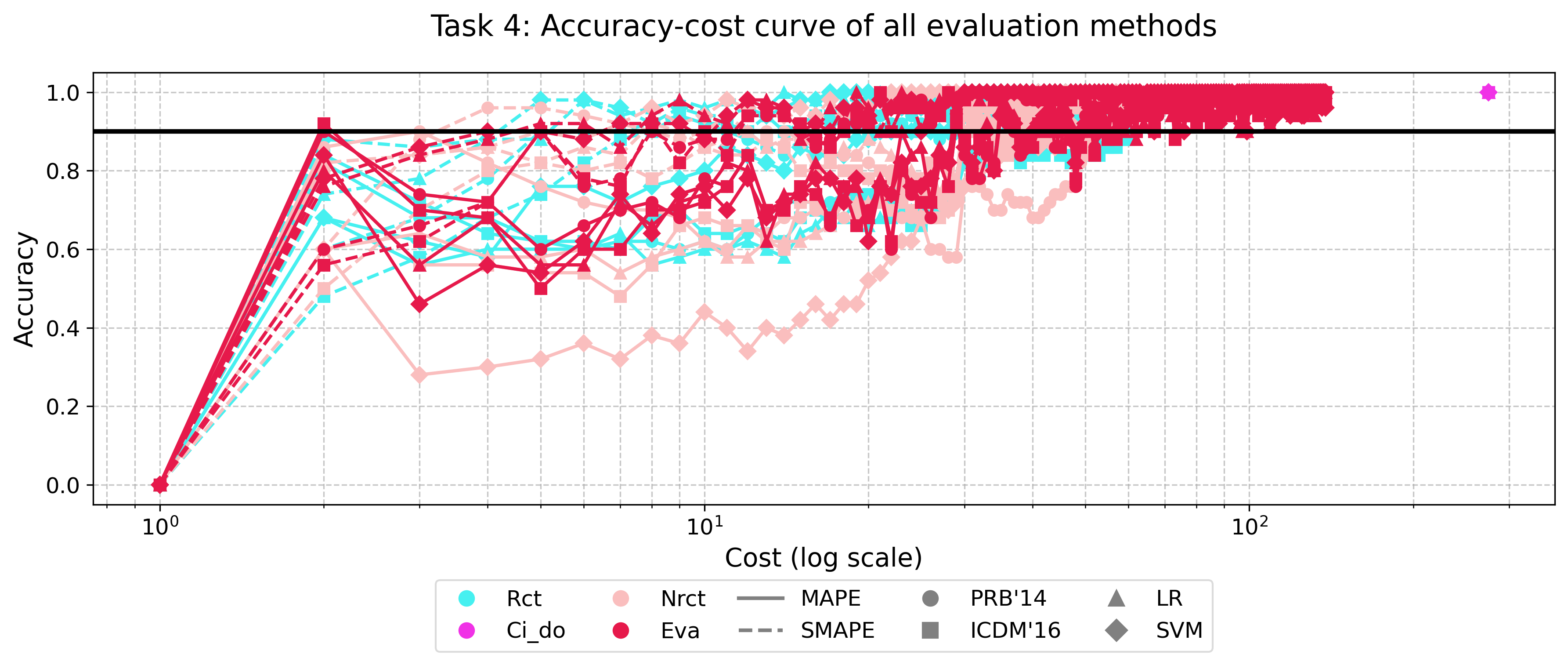}
    }
    \vspace{10pt}
    \subfigure[Task 5]{%
      \label{fig:task5}
      \centering
      \includegraphics[width=0.9\textwidth]{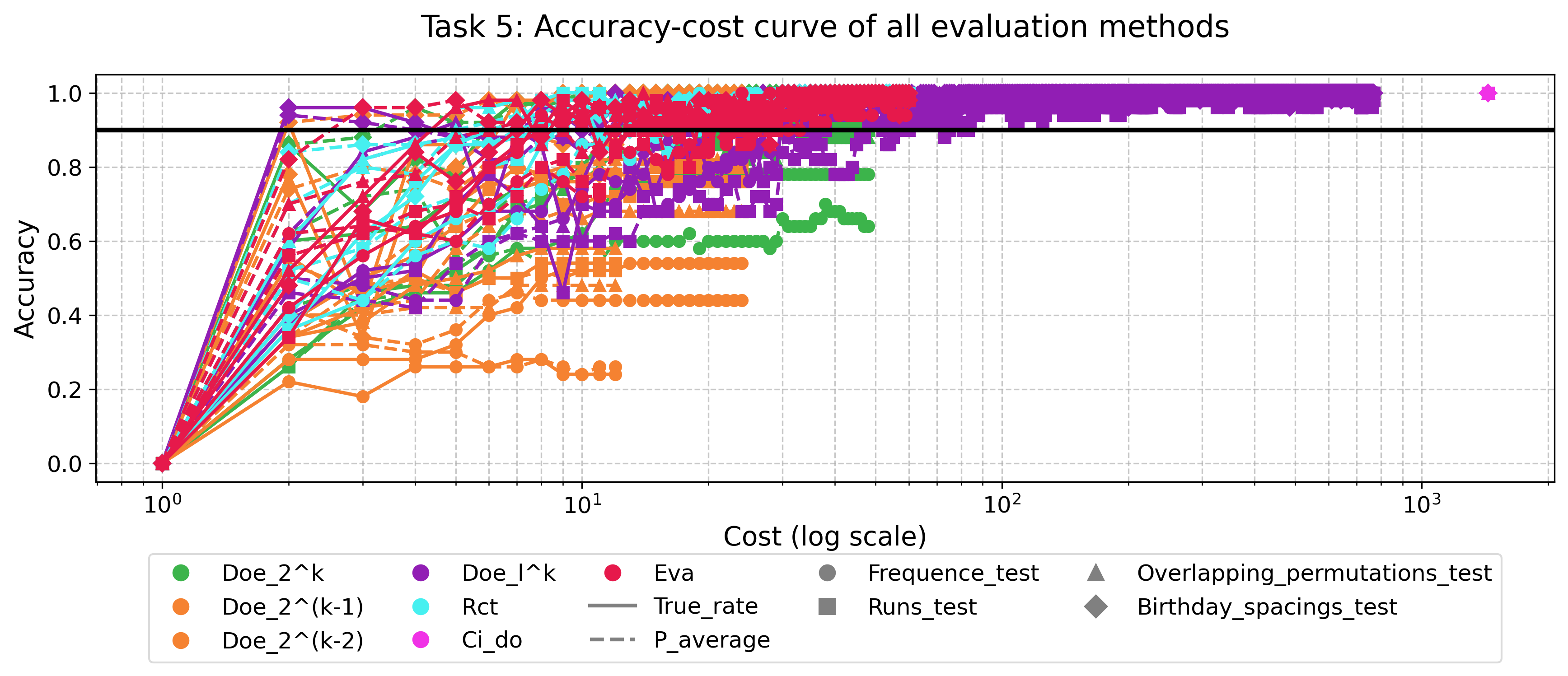}
    }
    \vspace{10pt}
    \subfigure[Task 6]{%
      \label{fig:task6}
      \centering
      \includegraphics[width=0.9\textwidth]{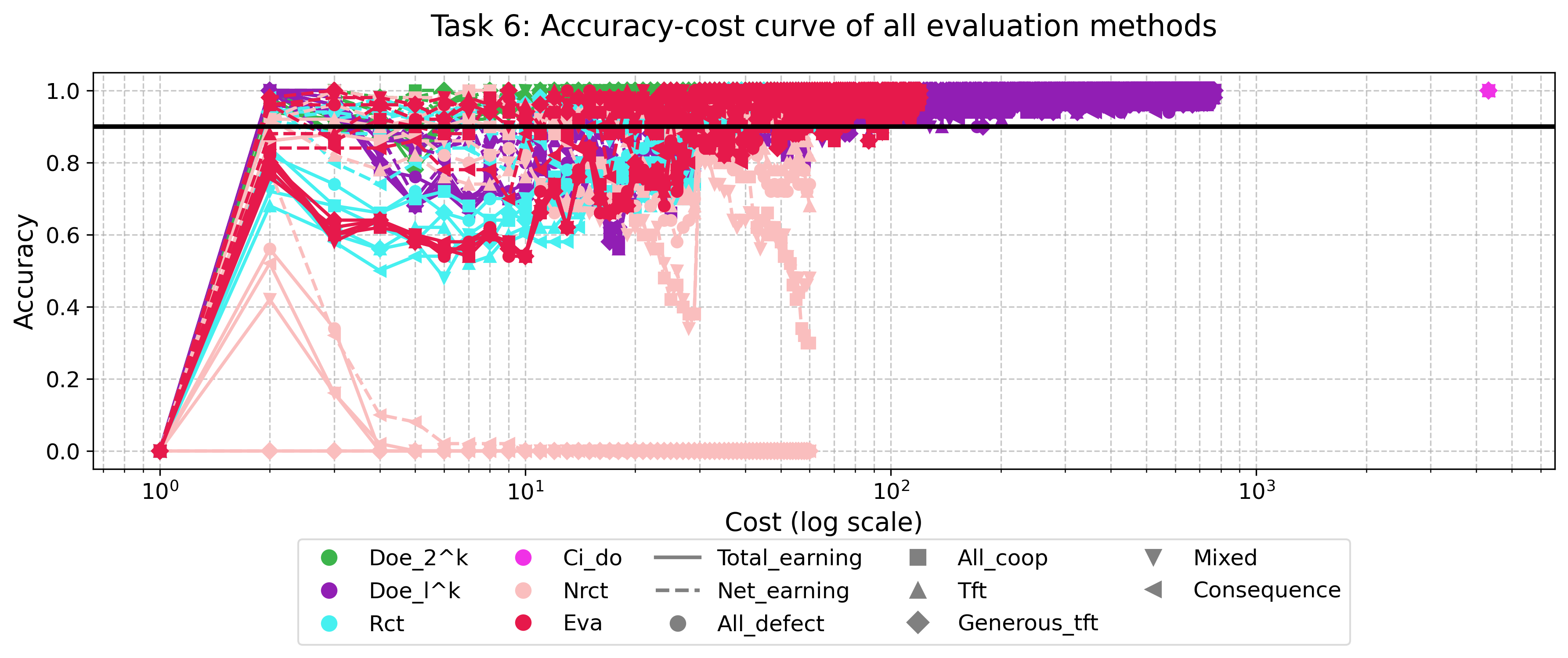}
    }
\caption{The accuracy–cost curves for various evaluation methods in Tasks 4, 5 and 6 are shown above. Colors represent different methods, shapes correspond to objects, and line styles denote evaluation metrics. The red lines indicate the Evaluatology method, which achieves the highest performance.}
\label{fig:task456}
\end{figure*}

In conclusion, both RCTs and evaluatology represent robust evaluation strategies for this task, with evaluatology offering superior cost efficiency. NRCTs may serve as a viable alternative where full randomization is impractical. Accuracy–cost curves for all methods are presented in Figure~\ref{fig:task4}.

\subsection{Pseudo-random number verification}

This task is defined as evaluating the ability of multiple randomness test algorithms of pseudo-random sequences, where the sequences are generated using the logistic map~\cite{may1976simple} with multiple initial values, parameters, and lengths, which is totally decisive rather than random. According to the definition, the less confident a randomness test algorithm holds for its randomness, the more correctness it achieves. We adopt four test algorithms from the Dieharder randomness testing tool~\cite{brown2018dieharder}: frequency test, runs test, overlapping permutations test, birthday spacings test. The evaluation indexes taken are the true rate that the sequence is randomly generated, and the average p-value measuring the confidence. This experiment aims at evaluating the indexes' accuracy estimation ability of different evaluation methods. The detailed illustration of this task is in Section~\ref{sec:methods}.

Our experimental results for this task are summarized as follows. First, the $2^k$ full-factorial design exhibited variable convergence in accuracy across the four test algorithms and the two performance indices: true rate and p-value. For the p-value index, maximum accuracies were 64\% for the frequency test, 90\% for the runs test, 88\% for the overlapping permutations test, and 100\% for the birthday spacings test. Corresponding values for the true rate index were 78\%, 92\%, 98\%, and 100\%, respectively. These results indicate that the $2^k$ method is effective for certain algorithms and metrics, but performance is inconsistent.

The $2^{k-p}$ design showed similar but reduced accuracy. For $p=1$, maximum p-value accuracies were 44\%, 76\%, 68\%, and 100\%, while true rate accuracies reached 54\%, 80\%, 82\%, and 98\%, respectively. For $p=2$, performance further declined to 26\%, 52\%, 48\%, and 78\% for the p-value index, and 24\%, 54\%, 58\%, and 86\% for the true rate index.

Second, in contrast, the $l^k$ DoE method with $l=4$ achieved accuracies exceeding 95\% across all algorithms and indices after approximately 100 evaluations, albeit at substantially higher cost.

Third, both RCTs and evaluatology methods attained over 95\% accuracy within 30 evaluations for both indices, barring minor out-of-distribution deviations. No further improvements were observed with additional evaluations. Performance was consistent across all four algorithms.

In conclusion, RCTs and evaluatology represent efficient and effective evaluation strategies for this task, with evaluatology offering marginally better cost efficiency. The $l^k$ DoE method is less desirable due to its high computational expense. Accuracy–cost curves for each method are presented in Figure~\ref{fig:task5}.

\subsection{Revenue evaluation of game strategies}

This task is defined as revenue evaluation of game strategies, measured by the effectiveness of the company's bidding strategy in multiple rounds of games. Unlike other tasks, this one is carried out in the form of multiple rounds of two-on-two competition. The evaluation conditions of the task include multiple values of game time, income matrix, decay factor, and noise, where the discount factor is defined as the ratio of the profit value of the next round to that of the current round, and the noise is defined as the probability of failure in making decisions. We defined the cumulative decay income as the evaluation index, fully explained in Section~\ref{sec:methods}. We evaluate the following six bidding strategies derived from the work of Axelrod et al.~\cite{axelrod1981evolution,axelrod1988further}: always defect, always cooperate, tic-for-tac, generous tic-for-tac, mixed strategy, consequence strategy. The evaluation indexes taken are total earning, and net earning on different occasions. This experiment aims at evaluating the indexes' accuracy estimation ability of different evaluation methods. The detailed illustration of this task is in Section~\ref{sec:methods}.

%Our experiment results of this task is as follows. First, the $2^k$ DoE method achieves over 95\% accuracy at the cost of 20 times of evaluations. This conclusion is the same for the predictions on every strategy. However, the $l^k$ DoE method taking $l=4$ has no accuracy improvement but a 16-times cost. Therefore, it is not a feasible evaluation method for this task. Second, the results for the RCTs method achieve over 90\% at the cost of 60 times of evaluation. The evaluatology method is slightly better compared to the RCTs method, due to the 5\% better accuracy performances. Third, the NRCTs method achieves bad performance. For the total earning index, the accuracies of six objects converge to 0\%, 96\%, 82\%, 0\%, 0\%, and 0\%. For the net earning index, they are 74\%, 30\%, 68\%, 0\%, 48\%, and 0\%. This means most NRCTs evaluations are impractical. As a conclusion, only the evaluatology method should be considered as the optimal method. Figure~\ref{fig:task6} shows the accuracy-cost curve of different evaluation methods in this task.

Our experimental findings for this task are summarized as follows. First, the $2^k$ full-factorial DoE method achieved the accuracy over 95\% across all strategic predictions within 20 evaluations. In contrast, an $l^k$ DoE configuration with $l=4$ yielded no improvement in accuracy despite a 16-fold increase in evaluation cost, rendering it impractical for this application.

Second, the RCTs method attained accuracies exceeding 90\% after 60 evaluations. The evaluatology method again demonstrated marginally superior performance, outperforming RCTs by approximately 5\% in final accuracy.

Third, non-randomized controlled trials (NRCTs) exhibited inconsistent and largely unsatisfactory results. For the total earning index, accuracies converged to 0\%, 96\%, 82\%, 0\%, 0\%, and 0\% across the six test objects. Corresponding values for the net earning index were 74\%, 30\%, 68\%, 0\%, 48\%, and 0\%, indicating that most NRCTs evaluations were ineffective or inapplicable.

In conclusion, the evaluatology method represents the optimal evaluation strategy for this task, balancing high accuracy with feasible computational cost. Accuracy–cost curves for each method are presented in Figure~\ref{fig:task6}.

\subsection{CPU performance evaluation}

This task is defined as evaluating CPU performances on a certain computation problem defined by SPEC CPU 2017~\cite{SPECCPU2017}. Two CPUs are involved in the experiment. Our experiment includes different evaluation methods that take the results on different numbers of copies/threads, dataset sizes, and compiler options. The only evaluation index is the running time. The detailed illustration of this task is in Section~\ref{sec:methods}.

%Our experiment results of this task is as follows. First, the $2^kr$ DoE method and the $l^kr$ method that $l=3$ and $r=3$ achieve 80\% accuracies at low cost of less than 30 times of evaluations. However, their accuracies both converge to 60\%, indicating their disability as evaluation methods for this task. This conclusion is the same for the performances on the two CPUs. Second, the RCTs method and the evaluatology method achieve over 95\% accuracy at the cost of 30 times of evaluations, if regardless of some minor out-of-distribution data, which have no further improvement possibility. As a conclusion, the RCTs method and the evaluatology method should be considered as the optimal method. Figure~\ref{fig:task7} shows the accuracy-cost curve of different evaluation methods in this task.

Our experimental results for this task are summarized as follows. First, both the $2^kr$ and $l^kr$ DoE methods with $l=3$, and $r=3$ achieved approximately 80\% accuracy at evaluation costs less than 30 evaluations. However, their accuracies subsequently plateaued near 60\%, indicating limited effectiveness as reliable evaluation strategies for this task. This trend was consistent across performance assessments on both two CPUs.

Second, the RCTs and evaluatology methods attained the accuracy over 95\% within 30 evaluations, barring minor deviations attributable to out-of-distribution samples. No further improvements were observed with additional evaluations.

\begin{figure*}
\centering
    \subfigure[Task 7]{%
      \label{fig:task7}
      \centering
      \includegraphics[width=0.9\textwidth]{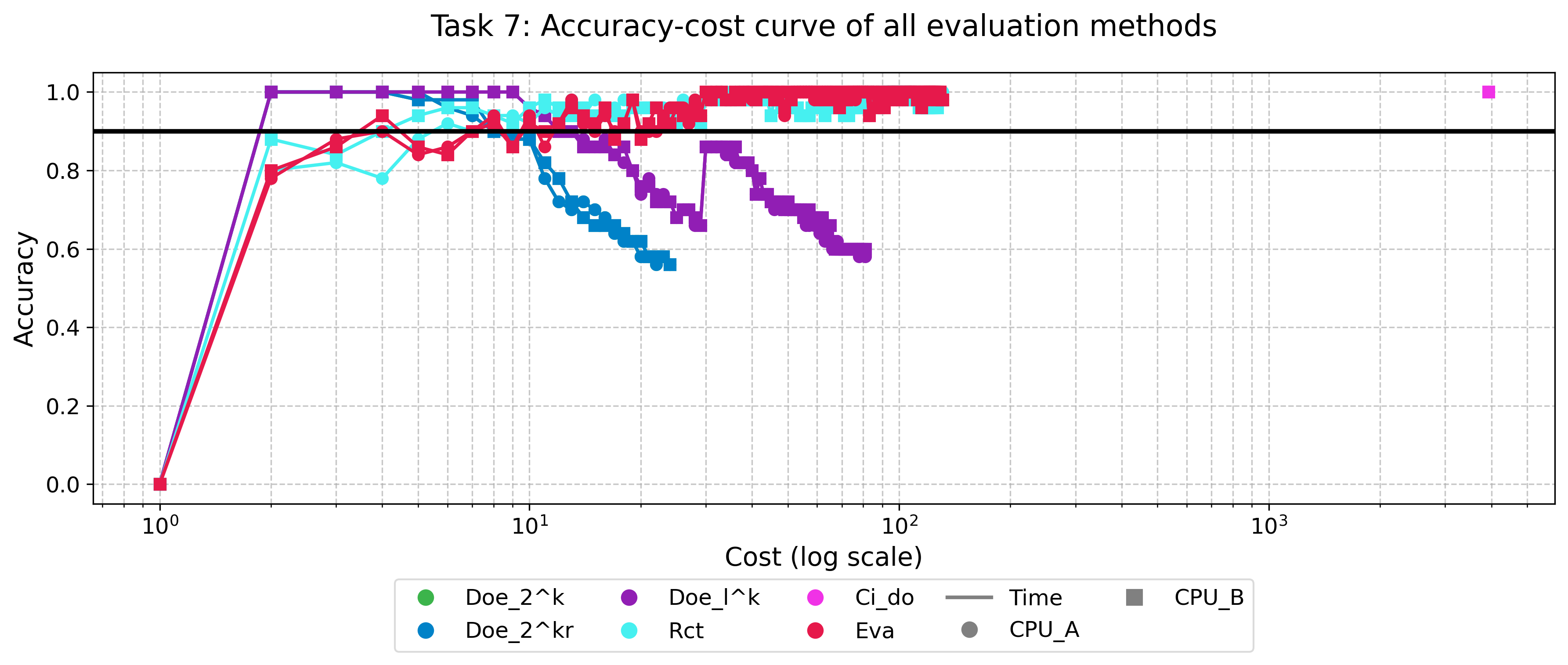}
    }
    \vspace{10pt}
    \subfigure[Task 8]{%
      \label{fig:task8}
      \centering
      \includegraphics[width=0.9\textwidth]{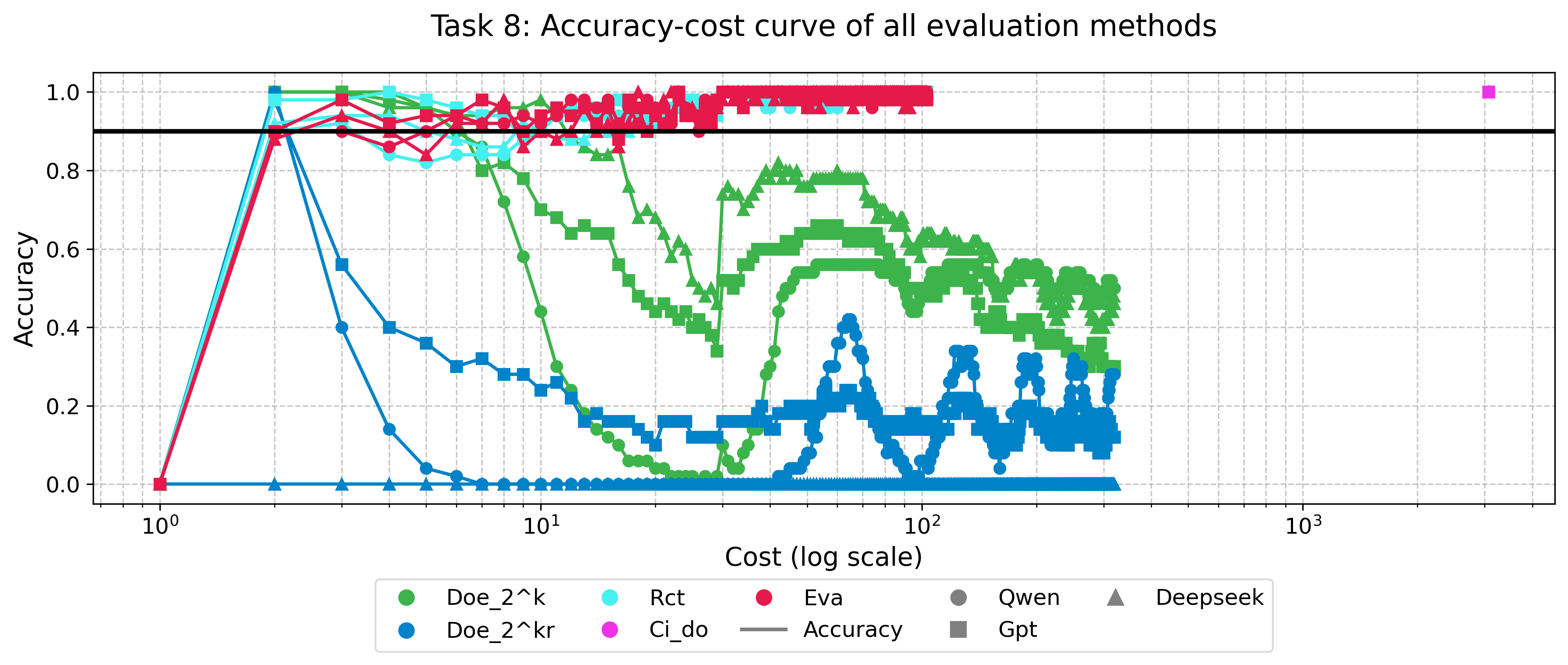}
    }
\caption{The accuracy–cost curves for various evaluation methods in Tasks 7 and 8 are shown above. Colors represent different methods, shapes correspond to objects, and line styles denote evaluation metrics. The red lines indicate the Evaluatology method, which achieves the highest performance.}
\label{fig:task78}
\end{figure*}

In conclusion, RCTs and evaluatology represent robust and efficient evaluation approaches for this task, demonstrating superior convergence and accuracy compared to DoE-based methods. Accuracy–cost curves for each method are presented in Figure~\ref{fig:task7}.

\subsection{large language model (LLM) problem solving}

This task is defined as evaluating the performance of given LLMs under various configurations of multiple-choice mathematical questions. All three LLMs evaluated are \texttt{gpt-4o-mini}~\cite{bubeck2023sparks}, \texttt{dschat}~\cite{liu2024deepseek}, and \texttt{qwen-2.5}~\cite{bai2023qwen}. Our experiment includes evaluation methods that check the accuracies of LLMs on 20 mathematical questions with given prompts: numbers of few-shot, chain of thoughts~\cite{kojima2022large}, choices permutation, language translation, choices modification, and capacity/role. The only evaluation index is the accuracy over all questions. The detailed illustration of this task is in Section~\ref{sec:methods}.

%Our experiment results of this task is as follows. First, the $2^k$ DoE method and the $2^kr$ method that $r=3$ have bad performances of accuracies and are not feasible evaluation methods for this task. Specifically, the accuracy of the $2^k$ method is almost never over 80\% and converges within 30\% to 50\% with enough the repetitions, while the accuracy of the $2^kr$ is almost within 0\% to 40\%, which is totally ineffective as an estimation. Second, the RCTs method and the evaluatology method achieve over 95\% accuracy at the cost of 30 times of evaluations, which have no further improvement possibility taking more cost. As a conclusion, the RCTs method and the evaluatology method should be considered as the optimal method for this task. Figure~\ref{fig:task8} shows the accuracy-cost curve of different evaluation methods in this task.

Our experimental findings for this task are summarized as follows. First, both the $2^k$ full-factorial design and the $2^kr$ repetition design with $r=3$ exhibited poor performance and are deemed unsuitable as evaluation methods. The $2^k$ method failed to exceed 80\% accuracy, with values converging between 30\% and 50\% upon sufficient repetition. The $2^kr$ design performed even more poorly, yielding accuracies consistently within 0\% and 40\%, confirming its ineffectiveness for reliable estimation.

Second, the RCTs and evaluatology methods achieved the accuracy over 95\% within 30 evaluations, with no further improvement observed upon increasing the evaluation budget.

In conclusion, RCTs and evaluatology represent the only acceptable evaluation methods for this task, demonstrating superior accuracy and efficiency. Accuracy–cost curves for all methods are presented in Figure~\ref{fig:task8}.

\subsection{Overall performances}

Table~\ref{tab:results} summarizes the overall performance of each evaluation method, assessed by accuracy, evaluation cost, and the Cost at 90\% Accuracy (C@0.9A), defined as the minimum number of evaluations required to achieve at least 90\% accuracy. A lower C@0.9A value indicates superior efficiency. If the method fails to reach 90\% accuracy, C@0.9A is defined as $\infty$.

Across most tasks, the evaluatology method~\cite{wang2024achieving} consistently demonstrated optimal performance, achieving the lowest C@0.9A values. The randomized controlled trials (RCTs) method ranked as the second most effective approach, offering robust accuracy at competitive evaluation costs. These results underscore the efficiency and reliability of evaluatology for high-accuracy evaluation under constrained cost concern.

\begin{table}[ht]
\caption{\textbf{Evaluation results}}\label{tab:results}
\begin{center}\begin{tabular}{lrcr}
\toprule
Methods           & Accuracy   & Cost   & C@0.9A  \\
\midrule
Obs               & 0$\sim$0.6 & high   & $\infty$  \\
Doe\_$2^k$        & 0.6$\sim$1 & low    & $\infty$  \\
Doe\_$2^kr$       & 0.2$\sim$1 & low    & $\infty$  \\
Doe\_$2^{k-p}$    & 0.2$\sim$1 & low    & $\infty$  \\
Doe\_$l^k$        & 0.6$\sim$1 & high   & $\infty$  \\
Rct               & 0.9$\sim$1 & low    & 30  \\
Ci\_do            & 1          & high   & 200+  \\
Ci\_scm           & 0$\sim$1   & high   & 200+  \\
Qe\_nrct          & 0.2$\sim$1 & middle & 60$\sim\infty$  \\
Qe\_stagger       & 0.9        & high   & 30$\sim$200+  \\
Eva               & 0.9$\sim$1 & low    & 30  \\
\bottomrule
\end{tabular}\end{center}
\footnotetext{This table presents the performance of various evaluation methods in meta-evaluation experiments.}
\end{table}

Our interpretation of the suboptimal performance exhibited by the observation method centers on its inherent reliance on the probability distribution of experimental condition (EC) values. In our simulations, which mirror real-world heterogeneity, certain EC values were assigned significantly higher occurrence probabilities -- differing by orders of magnitude -- than others. Consequently, as the number of observations increases, the aggregated results increasingly reflect the average behavior of high-probability instances rather than the true global average across the entire EC space. This sampling bias systematically skews the outcome and undermines the validity of observational estimates.

The inadequacy of DoE methods arises from two main limitations. First, DoE is fundamentally designed to elucidate the influence of various factors on experimental outcomes, not to isolate or quantify the causal effect of a specific object under study. Second, all DoE approaches rely on a sparse sampling of the configuration space, which is generally insufficient to capture the full complexity and variability of the ground-truth distribution. This insufficient sampling leads to significant estimation bias and weak correlation with results that would be obtained from exhaustive evaluation of the entire EC space.

In the task of LLM problem solving, we found cyclical fluctuation in performances of the $2^k$ and $2^kr$ DoE methods. Hereby, we provide probable explanation of this phenomenon. Taking the $2^k$ method as example, the EC space of this task is shaped $8\times4\times4\times3\times2\times4=3072$, while that of the size of the $2^k$ DoE method is $2^6=64$. As a result, different configurations are chosen in every 64 times of evaluation. If less than 64, there must be some configurations that have some unique characteristics that are not chosen. Therefore, the performances close to the integer times of 64 are better, as they are more close to the global average performance of the entire EC space.

The outcome of the intervention staggering method in Task 2 shows a somewhat yet not critical accuracy decrease compared to the controlled experiments and evaluatology method. This is because the rainfall prediction task in now-casting scenario is not sensitive to the non-recent data.

There is a leap of accuracy at about 30 evaluation times cost, because the distribution estimating methods are different. Specifically, Z-score less than 30 and T-score otherwise.

\section{Discussion}\label{sec:discussion}

In this work, we advance meta-evaluation from a largely conceptual notion to an operational scientific framework. By defining the evaluation space, formalizing its representation, and instantiating it through AxiaBench, we enable -- for the first time -- a systematic, quantitative comparison of ten representative evaluation methods across eight canonical scientific domains.

Our analyses demonstrate that no existing method simultaneously delivers accuracy and efficiency across diverse scenarios, except a very recently proposed entire-space stratified sampling method~\cite{wang2024achieving} based on evaluatology~\cite{zhan2024evaluatology} consistently achieves superior performance with substantially lower cost. Together, these results position meta-evaluation not merely as a methodological exercise, but as an essential instrument for trustworthy scientific inquiry.

Several limitations merit consideration. First, the effectiveness of our approach diminishes in settings where the evaluation condition space is intrinsically small, poorly characterized, or inaccessible, as stratification relies on a sufficiently rich configuration space. Second, while our benchmark spans a wide range of domains, some tasks necessarily rely on simulated rather than real-world experimental data, raising questions about external validity. We mitigated this trade-off by carefully balancing realism with tractability, and by cross-checking simulated findings against domains where ground-truth data are available. Nonetheless, extending AxiaBench to more real-world experimental pipelines remains an important direction for future work.

%As a limitation, this work and the evaluatology method does not work in good performance when the EC space is small or inaccessible. Another limitation is that the external effectiveness of this work may be questioned, because part of the evaluation designs are simulation instead of real-world experiments. As an explanation, we spared no effort in the balance between restoring real-world scenarios and experiment cost. We argue that there are no or few influence on the effectiveness of our novel evaluation methodology, either using real-world data achieved or \textit{ab initio} computed or simulated data.

Despite these limitations, the broader implications of this work are profound. By elevating meta-evaluation to a formal scientific object, we provide both a conceptual foundation and a practical tool set for interrogating the reliability of evaluation itself -- a question that underpins all empirical disciplines. AxiaBench enables researchers to uncover when and why established evaluation methods fail, and offers a pathway toward principled, generalizable alternatives. Beyond methodological refinement, this work contributes to a larger scientific vision: bridging the gap between experimental designs and the complexity of real-world systems, accelerating the deployment of evaluation methods that are both rigorous and efficient, and ultimately strengthening the epistemic foundations on which scientific progress depends.

%Through estimating the ground-truth evaluation outcomes of different objects, our meta evaluation methodology and AxiaBench provide a systematic approach for measuring and analyzing the gap between the outputs of AI models and the ground-truth data derived from the real world. In addition, the analysis of evaluation results of different objects in multiple configurations provide causal explanation of the effect of objects. This work lays a solid foundation for bridging the gap between experimental settings and a high dimensional space of the real-world applications, not only accelerating the development and deployment of the novel evaluation methods, but also affirming the usability of existing evaluation methods (e.g., RCTs, and DoE methods in partial tasks) in various fields, which can bring about efficiency improvements that have not been thoroughly studied before this work.

\section{Methods}\label{sec:methods}

%This section introduces the entire evaluation experiments designs and configurations of celestial motion prediction~\cite{lichtenberg2013regular,cambel1993applied}, rainfall prediction~\cite{burrows1991objective}, population dynamics simulation~\cite{turing1990chemical,tilman1997spatial}, formation energy prediction~\cite{wicks1963thermodynamic,kohn1965self}, pseudo-random number verification~\cite{may1976simple,rukhin2001statistical}, revenue evaluation of game strategies~\cite{axelrod1981evolution}, CPU performance evaluation~\cite{wang2024achieving}, and LLM problem solving~\cite{yang2023gpt,liu2024finemath} on observation, $2^k$, $2^{k-p}$, $2^kr$, $l^k$~\cite{jain1990art}, RCTs~\cite{chalmers1981method}, do-calculus~\cite{imbens2015causal,powell2018book}, SCM~\cite{imbens2015causal,powell2018book}, NRCTs~\cite{thyer2012quasi}, intervention staggering method~\cite{thyer2012quasi}, and evaluatology method. Designs, configurations, and evaluated methods are slightly different among the eight tasks, and the detailed description is shown in Table~\ref{tab:designs}. We specifically explain how the result of every task is computed, and how the accuracy and cost of every evaluation method is computed in this section. Since most tasks (except Task 7 and Task 8) have no randomness at all after fixing the seeds (0, 37, 42 are actual settings in this work), the cost of repeating experiments can be reduced. Figure~\ref{fig:methods} is the workflow and design of this work.

\begin{figure}
	\centering
		\includegraphics[scale=.4]{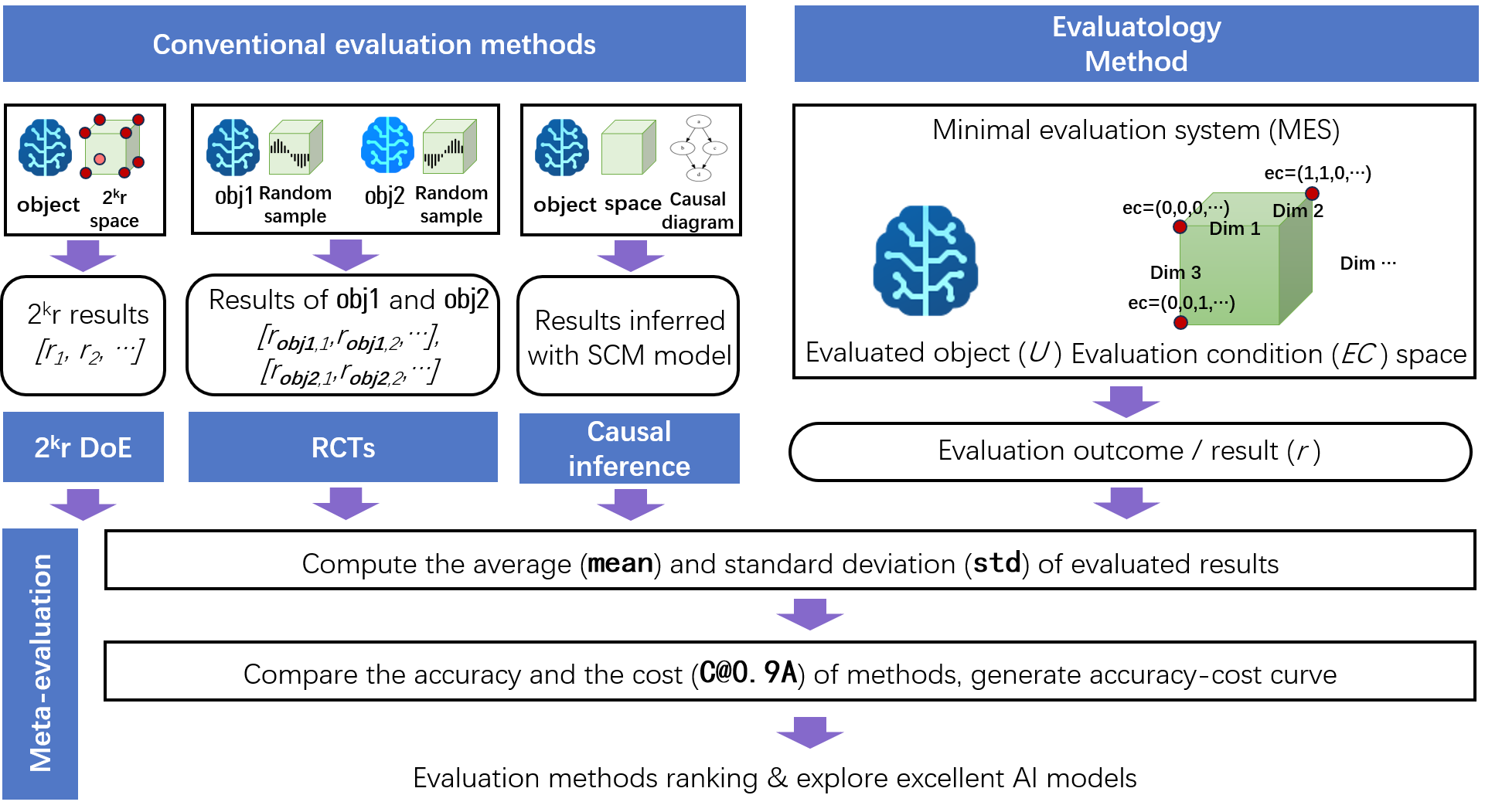}
	\caption{The process and structure of meta-evaluation and its implementation on assessment methodologies, encompassing our innovative evaluatology approach.}
	\label{fig:methods}
\end{figure}

This section outlines the experimental design and configurations for evaluating a range of computational tasks, including celestial motion prediction~\cite{lichtenberg2013regular,cambel1993applied}, rainfall prediction~\cite{burrows1991objective}, population dynamics simulation~\cite{turing1990chemical,tilman1997spatial}, formation energy prediction~\cite{wicks1963thermodynamic,kohn1965self}, pseudo-random number verification~\cite{may1976simple,rukhin2001statistical}, revenue evaluation of game strategies~\cite{axelrod1981evolution}, CPU performance evaluation~\cite{wang2024achieving}, and large language model (LLM) problem-solving~\cite{yang2023gpt,liu2024finemath}. Each task is assessed using multiple evaluation methodologies: observation, DoE designs ($2^k$, $2^{k-p}$, $2^kr$, $l^k$)~\cite{jain1990art}, randomized controlled trials (RCTs)~\cite{chalmers1981method}, do-calculus~\cite{imbens2015causal,powell2018book}, structural causal model (SCM)~\cite{imbens2015causal,powell2018book}, non-randomized controlled trials (NRCTs)~\cite{thyer2012quasi}, intervention staggering~\cite{thyer2012quasi}, and the evaluatology method~\cite{wang2024achieving}. Experimental designs, parameter configurations, and evaluation methods vary across the eight tasks, as detailed in Table~\ref{tab:designs}. We provide explicit descriptions of how results are generated for each task, and how accuracy and evaluation cost are quantified for each method. To ensure reproducibility and minimize stochastic variability, all tasks (with the exception of Task 7 and Task 8) were executed using fixed random seeds (0, 37, and 42), thereby reducing the need for repeated trials. The overall workflow and experimental architecture are illustrated in Figure~\ref{fig:methods}.

\begin{table}[ht]
\caption{\textbf{Evaluation designs}}\label{tab:designs}
\resizebox{\linewidth}{!}{\begin{tabular}{lcccccccc}
\toprule
Methods         & Task 1 & Task 2 & Task 3 & Task 4 & Task 5 & Task 6 & Task 7 & Task 8  \\
\midrule
Obs             & Done   & Done   & Done   & N/A    & N/A    & N/A    & N/A    & N/A     \\
Doe\_2\textasciicircum k      & Done   & Done   & Done   & N/A    & Done   & Done   & Done   & Done    \\
Doe\_2\textasciicircum kr     & N/A    & N/A    & N/A    & N/A    & N/A    & N/A    & Done   & Done    \\
Doe\_2\textasciicircum(k-p)  & N/A    & N/A    & Done   & N/A    & Done   & N/A    & N/A    & N/A     \\
Doe\_l\textasciicircum k      & Done   & N/A    & Done   & N/A    & Done   & Done   & Done   & N/A     \\
Rct             & Done   & Done   & Done   & Done   & Done   & Done   & Done   & Done    \\
Ci\_do          & Done   & Done   & Done   & Done   & Done   & Done   & Done   & Done    \\
Ci\_scm         & Done   & Done   & N/A    & N/A    & N/A    & N/A    & N/A    & N/A     \\
Qe\_nrct        & N/A    & N/A    & N/A    & Done   & N/A    & Done   & N/A    & N/A     \\
Qe\_stagger     & N/A    & Done   & Done   & N/A    & N/A    & N/A    & N/A    & N/A     \\
Eva             & Done   & Done   & Done   & Done   & Done   & Done   & Done   & Done    \\
\bottomrule
\end{tabular}}
\footnotetext{This table provides a detailed account of the evaluated methods for each task, with ``Done'' indicating conducted and analyzed experiments, and ``N/A'' denoting experiment designs that are not applicable. The rationale behind these distinctions is elucidated in Section ~\ref{sec:results}. The abbreviations of the methods are clarified in preceding sections.}
\end{table}

\subsection{Celestial motion prediction}

As defined in Section~\ref{sec:results}, this task evaluates accuracies and costs of different evaluation methods. The methods evaluates the degree of chaos of given Lorenz system equations. In formal definition, a stranger attractor of a certain Lorenz system requires three parameters $a,b,c$ and the initial values $x,y,z$. The full form of a system is as follows:\\
\centerline{$\begin{cases}
\dfrac{\text{d}x}{\text{d}t}=f(x,y,z)\\
\dfrac{\text{d}y}{\text{d}t}=g(x,y,z)\\
\dfrac{\text{d}z}{\text{d}t}=h(x,y,z)
\end{cases}$,}\\
where $f,g,h$ are usually linear or pseudo-linear functions. This task simulates equations with precision $p$ for $N$ steps, generating sequences. The Lyapunov exponent~\cite{dingwell2006lyapunov} and Kolmogorov-Smirnov (KS) statistics~\cite{fasano1987multidimensional} (3-dimensional) are used to measure the chaos degree of sequences.

The EC space is defined as $p\times(a,b,c)\times N\times(x,y,z)$ with given values as follows:\\
\[
\begin{aligned}
&p \in \{0.01,\ 0.005,\ 0.001,\ 0.0005,\ 0.0001\} \\
&(a,b,c) \in\left\{ \begin{aligned} 
(0.1, 0.2, 5.0),\ (0.5, 1.2, 12.0),\ (1.0, 0.3, 20.0),\ (1.5, 1.8, 8.0),\\ (0.3, 1.6, 30.0), \ (1.8, 0.6, 18.0),\ (0.7, 1.1, 25.0),\ (1.3, 0.9, 35.0),\\
(0.2, 1.7, 10.0),\ (1.9, 0.4, 40.0)
\end{aligned} \right\} \\
&N \in \{200,\ 400,\ 800,\ 1600\} \\
&(x,y,z) \in\left\{ \begin{aligned} 
(1.0, 5.0, 10.0),\ (-2.0, -8.0, -15.0),\ (3.0, -12.0, 20.0),\\
(-4.0, 16.0, -25.0),\ (5.0, 20.0, 30.0), \ (-6.0, -24.0, -35.0),\\
(7.0, -28.0, 40.0),\ (-8.0, 32.0, -45.0),\ (9.0, 36.0, 50.0),\\
(-10.0, -40.0, -55.0)
\end{aligned} \right\}
\end{aligned}
\]
For a certain group of $(p,(a,b,c),N,(x,y,z))$ values, the sequence is generated by the above definition. In total, there are $5\times10\times4\times10=2000$ number of configurations.

The evaluated objects are Lorentz attractor~\cite{lorenz2017deterministic} (\texttt{obj1}), and R\"ossler attractor~\cite{rossler1983chaotic} (\texttt{obj2}), defined as follows:\\
Lorentz attractor: $\begin{cases}
\dfrac{\text{d}x}{\text{d}t}=a(y-x)\\
\dfrac{\text{d}y}{\text{d}t}=x(b-z)-y\\
\dfrac{\text{d}z}{\text{d}t}=xy-cz
\end{cases}$,
R\"ossler attractor: $\begin{cases}
\dfrac{\text{d}x}{\text{d}t}=-y-z\\
\dfrac{\text{d}y}{\text{d}t}=x+ay\\
\dfrac{\text{d}z}{\text{d}t}=b+z(x-c)
\end{cases}$.\\
Each object generate a sequence with the above configuration of parameters. The Lyapunov exponent and the KS statistics are indexes, computed with the sequence. So far, the meta evaluation result of every index and every object is defined and computed.

Seven evaluation methods (observation, $2^k$, $l^k$, RCTs, do-calculus, SCM, and evaluatology methods) compute with the result of every index and every object as the input, and the accuracies and costs of evaluation as the output, as shown in Section~\ref{sec:results}. The detailed computing method is explained in this section right after the explanation of eight tasks.

\subsection{Rainfall prediction}

As defined in Section~\ref{sec:results}, this task evaluates accuracies and costs of different evaluation methods. The methods evaluates the accuracy of the regression models in prediction on the \textit{ab initio} precipitation sequences. In formal definition, a sequence $R(t)$ is given by\\
\centerline{$R(t)=\rho\dfrac{q_S(1+\epsilon(t))L\Gamma_mw(1+\sigma_W\eta(t))}{R_\nu (T+\sigma_T\xi(t))^2}+\sigma_R\nu(t)$,}\\
where $t$ is the time in a day referring the corresponding sequence position, $\rho\equiv3.6$ is the air density, $q_S$ is the saturation specific humidity, $\epsilon$ is the vapor noise, $L\equiv2.5\times10^6$ is the latent heat of vaporization, $\Gamma_m\equiv0.006$ is the moist adiabatic lapse rate, $w$ is the updraft airflow speed, $\sigma_W$ is the airflow noise coefficient, $\eta$ is the airflow noise, $R_\nu\equiv461$ is the water gas constant, $T$ is the temperature, $\sigma_T$ is the temperature noise coefficient, $\xi$ is the temperature noise, $\sigma_R$ is the observation noise strength, and $\nu$ is the observation noise. All the noises are in normal distribution. The range of $t$ is within $[0,8760)$ representing the precipitation conditions per hour in an entire year.

The EC space is defined as $q_S\times w\times T,\times\sigma_W\times\sigma_T\times\sigma_R$ with given values as follows:\\
\[
\begin{aligned}
&q_S \in \{0.01,\ 0.02,\ 0.04\} \\
&w \in \{0.01,\ 0.1\} \\
&T \in \{253,\ 261,\ 269,\ 277,\ 285,\ 293,\ 301,\ 309,\ 317,\ 325\} \\
&\sigma_W \in \{0.01,\ 0.03\} \\
&\sigma_T \in \{0.5,\ 1,\ 1.5,\ 2\} \\
&\sigma_R \in \{0.1,\ 0.2,\ 0.3,\ 0.4,\ 0.5\}
\end{aligned}
\]
For a certain group of $(q_S,w,T,\sigma_W,\sigma_T,\sigma_R)$ values, the precipitation sequence is generated by the above definition. In total, there are $3\times2\times10\times2\times4\times5=2400$ number of sequences.

The evaluated objects are four classical regression models~\cite{zuur2009glm}: GLM (\texttt{obj1}), GAM (\texttt{obj2}), RF (\texttt{obj3}), and FBM (\texttt{obj4}). The accuracy of each object on each sequence is computed according to indexes of MSE, RMSE, and MAE. So far, the meta evaluation result of the index and objects is decided and obtained.

Seven evaluation methods (observation, $2^k$, RCTs, do-calculus, SCM, intervention staggering, and evaluatology methods) compute with the result of every index and every object as the input, and the accuracies and costs of evaluation as the output, as shown in Section~\ref{sec:results}. The detailed computing method is explained in this section right after the explanation of eight tasks.

\subsection{Population dynamics simulation}

As defined in Section~\ref{sec:results}, this task evaluates accuracies and costs of different evaluation methods. The methods evaluates population dynamic stability of different species. This task simulates the three species based on cellular automata in a two-dimensional grid, with states of the cells and rules defined above. In formal definition, a grid is a $a\times a$ sized matrix with positive integer values that stands for the amount of species. The initial distribution of species is decided by $A_{init}$ that has six patterns with different centers and non-centers. This task simulates the species' evolution with grids for $n$ steps, with given environment capacity $L$ and resource density $s$. According to the real-world condition, the environment capacity provides a constraint of the upper bound of species density. The rank of environment capacity sensitivity is: bacteria>yeast>algae (This task sets $24,11,7$ as coefficients  respectively). The resource density influences the evolution speed and sustainability. The rank of resource density demand is: yeast>algae>bacteria (This task sets $0.6,0.5,0.3$ as coefficients respectively).

The EC space is defined as $a\times A_{init}\times n\times L\times s$ with given values as follows:\\
\[
\begin{aligned}
&a \in \{10, 20, 50, 100\} \\
&A_{\text{init}} \in \left\{ \begin{aligned} \text{up centered}, \text{down centered}, \text{left centered},\\
\text{right centered}, \text{centered}, \text{spread}\end{aligned}\right\} \\
&n \in \{5, 10, 25, 100\} \\
&L \in \{4, 5, 6, 7\} \\
&s \in \{1.0, 1.5, 2.0, 2.5\}
\end{aligned}
\]
For a certain group of $(a,A_{init},n,L,s)$ values, the cellular grid is generated by the above definition. In total, there are $4\times5\times4\times4\times4=1536$ number of configurations.

The evaluated objects are bacteria~\cite{ben2000cooperative} (\texttt{obj1}), yeast~\cite{pfeiffer2001cooperation} (\texttt{obj2}), and algae~\cite{huisman1999biodiversity} (\texttt{obj3}). The stability of each object on each cellular grid is computed according to four indexes as follows: average population, standard deviation of the population, half-life, and growth rate~\cite{murray2007mathematical}. So far, the meta evaluation result of the index and objects is decided and obtained.

Eight evaluation methods (observation, $2^k$, $2^{k-p}$, $l^k$, RCTs, do-calculus, intervention staggering, and evaluatology methods) compute with the result of every index and every object as the input, and the accuracies and costs of evaluation as the output, as shown in Section~\ref{sec:results}. The detailed computing method is explained in this section right after the explanation of eight tasks.

\subsection{Formation energy prediction}

As defined in Section~\ref{sec:results}, this task evaluates accuracies and costs of different evaluation methods. The methods evaluates the performance of material formation energy prediction models. A set of materials are given with their structure formulae and formation energy value computed with \textit{ab initio} methods within a certain range of errors. The materials are defined by four general molecule modules $M$, and atom/ion types $X,Y,Z$. The formation energy $\Delta H_f$ of a material is computed based on $\Delta H_f=\sum_in_iE_{atom,i}-\sum_jk_jE_{bond,j}$, where $n_i$ and $k_j$ are the numbers of the $i$-th kind of atom and the $j$-th kind of chemical bond respectively, while $E_{atom,i}$ and $E_{bond,j}$ are formation energies of the $i$-th kind of elementary substance and $j$-th kind of chemical bond respectively, given as chemical constants in Table~\ref{tab:energy}.

\begin{table}[ht]
\caption{\textbf{Formation energy (kJ/mol)}}\label{tab:energy}
\begin{tabular}{cccccccc}
\ce{Na+} & 0.0 & \ce{-H} & 218.0 & \ce{-OH} & 4.2 & \ce{-CH3} & 146.7 \\
\ce{-NH2} & 190.0 & \ce{-C#N} & 302.4 & \ce{=O} & 249.2 & \ce{=CH2} & 385.0 \\
\ce{=NH} & 260.0 & \ce{=C=O} & 749.0 & \ce{H-H} & -436.0 & \ce{O-H} & -463.0 \\
\ce{C-H} & -413.0 & \ce{N-H} & -391.0 & \ce{C-C} & -348.0 & \ce{C-O} & -358.0 \\
\ce{C-N} & -305.0 & \ce{O-O} & -142.0 & \ce{O-N} & -200.0 & \ce{N-N} & -163.0 \\
\ce{Na-H} & -188.7 & \ce{Na-O} & -260.0 & \ce{Na-C} & -150.0 & \ce{Na-N} & -243.0
\end{tabular}
\footnotetext{The formation energy of substances and chemical bonds.}
\end{table}

The EC space is defined as $M\times X\times Y\times Z$ with given values as follows:\\
\[
\begin{aligned}
&M \in \{\text{X}_2, \text{XYZ}, \text{XZ}, \text{XY}_2\text{Z}\} \\
&X \in \{\ce{Na+}, \ce{-H}, \ce{-OH}, \ce{-CH3}, \ce{-NH2}, \ce{-}\chemfig{C\equiv N}\} \\
&Y \in \{\ce{-O-}, \ce{-CH2-}, \ce{-NH-}, \ce{-}\chemfig{C(=[2,0.5]O)}\ce{-}\} \\
&Z \in \{\ce{Na+}, \ce{-H}, \ce{-OH}, \ce{-CH3}, \ce{-NH2}, \ce{-}\chemfig{C\equiv N}\}
\end{aligned}
\]
Unlike other tasks, there are $5+6\times5\times4+6\times5+6\times5\times4=275$ number of substances in total, due to the duplication removal and illegal chemical formulae.

The evaluated objects are: PRB'14~\cite{meredig2014combinatorial} (\texttt{obj1}), ICDM'16~\cite{agrawal2016formation} (\texttt{obj2}), LR (\texttt{obj3}), and SVM (\texttt{obj4}). Each object inputs the substances and outputs its formation energy predicted, evaluated by MAPE and SMAPE. So far, the meta evaluation result of every index and every object is defined and computed.

Four evaluation methods (RCTs, do-calculus, NRCTs, and evaluatology methods) compute with the result of every index and every object as the input, and the accuracies and costs of evaluation as the output, as shown in Section~\ref{sec:results}. The detailed computing method is explained in this section right after the explanation of eight tasks.

\subsection{Pseudo-random number verification}

As defined in Section~\ref{sec:results}, this task evaluates accuracies and costs of different evaluation methods. The methods evaluates the ability of multiple randomness test algorithms on certain logistic map~\cite{may1976simple} sequences. In formal definition, a sequence $y_n$ is given by\\
\centerline{$x_{n+1}=rx_n(1-x_n),y_n=\sum_{i=1,\cdots,N}x_{n,i},\text{ for any }n\in|y|$,}\\
where $|y|$ stands for the sequence length, $x_{n,i}$ is the $i$-th logistic map, $r$ is the mapping coefficient, $X_0=x_{1,0},\cdots,x_{n,0}$ are initial values, and $N$ is the number of maps.

The EC space is defined as $N\times X_0\times r\times|y|$ with given values as follows:\\
\[
\begin{aligned}
&N \in \{1, 2, 3, 4, 5, 6, 7, 8, 9, 10\} \\
&X_0 \in \{0.1, 0.2, 0.3, 0.4, 0.5, 0.6, 0.7, 0.8, 0.9\} \\
&r \in \{-1.8, -1.5, 3.6, 3.8\} \\
&|y| \in \{10, 20, 50, 100\}
\end{aligned}
\]
For a certain group of $(N,X_0,r,|y|)$ values, the sequence $y_n$ is generated by the above definition. In total, there are $10\times9\times4\times4=1440$ number of sequences.

The evaluated objects are defined as frequency test (\texttt{obj1}), runs test (\texttt{obj2}), overlapping permutations test (\texttt{obj3}), birthday spacings test (\texttt{obj4}), derived from the Dieharder tool~\cite{brown2018dieharder}. Each object runs and output the confidence that the sequence is randomized, given by the two indexes: the true rate, and the average p-value as defined in Section~\ref{sec:results}. So far, the meta evaluation result of every index and every object is defined and computed.

Five evaluation methods ($2^k$, $l^k$, RCTs, do-calculus, and evaluatology methods) compute with the result of every index and every object as the input, and the accuracies and costs of evaluation as the output, as shown in Section~\ref{sec:results}. The detailed computing method is explained in this section right after the explanation of eight tasks.

\subsection{Revenue evaluation of game strategies}

As defined in Section~\ref{sec:results}, this task evaluates accuracies and costs of different evaluation methods. The methods evaluates the effectiveness of the company's bidding strategy in multiple rounds of games. In formal definition, a company has to decide its strategy in $n$ rounds. In each round, there are two choices: make a discount, do not make a discount. There is a probability $\epsilon$ (the noise coefficient defined in Section~\ref{sec:results}) for a company to execute a wrong decision. The income matrix for two companies is as follows:\\
\centerline{$\begin{matrix}
                       & \text{Com. B disc.} & \text{Com. B no disc.}\\
\text{Com. A \quad~disc.}    & (b,b)               & (d,a)\\
\text{Com. A no disc.} & (a,d)               & (c,c)\\
\end{matrix}$}\\
In the above matrix, $a,b,c,d$ are coefficients with restriction $a<b<c<d$. The total earning $U$ of a company after $n$ rounds of gaming is computed as: $U=\sum_{i=1}^\infty\delta^{i-1}v(i)$, where $v(i)$ is the earning at the $i$-th round, $\delta>0.5$ is the decay factor defined in Section~\ref{sec:results}. The computation process of $U$ is actually a finite summation, due to the finite round number $n$.

The EC space is defined as $n\times(a,b,c,d)\times\delta\times\epsilon$ with given values as follows:\\
\[
\begin{aligned}
&n \in \{2, 5, 10, 100\} \\
&(a,b,c,d) \in 
\left\{ \begin{aligned} 
(2,4,5,6),\ (1,4,5,6),\ (2,4,6,7),\ (1,4,6,7),\ (2,3,5,7), \\
(1,3,5,7),\ (2,3,6,7),\ (1,3,6,7),\ (2,5,6,7),\ (1,5,6,7)
\end{aligned} \right\} \\
&\delta \in \{0.5, 0.58, 0.66, 0.74, 0.82, 0.9, 0.95, 0.975, 0.99\} \\
&\epsilon \in \{0.0, 0.05, 0.1, 0.15, 0.2, 0.25, 0.3, 0.35, 0.4, 0.45, 0.5, 1.0\}
\end{aligned}
\]
For a certain group of $(n,(a,b,c,d),\delta,\epsilon)$ values, the game procedure is decided by the above definition. In total, there are $4\times10\times9\times12=4320$ number of configurations.

The evaluated objects are six bidding strategies derived from the work of Axelrod et al.~\cite{axelrod1981evolution,axelrod1988further}: always defect (\texttt{obj1}), always cooperate (\texttt{obj2}), tic-for-tac (\texttt{obj3}), generous tic-for-tac (\texttt{obj4}), mixed strategy (\texttt{obj5}), and consequence strategy (\texttt{obj6}). Each object competes with other objects and output its earnings, given by the two indexes: total earning, and net earning as defined in Section~\ref{sec:results}. So far, the meta evaluation result of every index and every object is defined and computed.

Five evaluation methods ($2^k$, $l^k$, RCTs, do-calculus, and evaluatology methods) compute with the result of every index and every object as the input, and the accuracies and costs of evaluation as the output, as shown in Section~\ref{sec:results}. The detailed computing method is explained in this section right after the explanation of eight tasks.

\subsection{CPU performance evaluation}

As defined in Section~\ref{sec:results}, this task evaluates accuracies and costs of different evaluation methods. The methods evaluates CPU performances on a certain computation problem defined by SPEC CPU 2017~\cite{SPECCPU2017}. The problem is defined as a recursive solution generator for Sudoku, noted as \texttt{648.exchange2\_s} in the SPEC intspeed benchmark suite. The workload is run with multiple choices of threads numbers $th$, compiler options $c$, and Sudoku numbers $size$.

The EC space is defined as $th\times c\times size$ with given values as follows:\\
\[
\begin{aligned}
&th \in \{1, 2, 3, \cdots, 127, 128, 200, 256, 300\} \\
&c \in \{\text{O1}, \text{O2}, \text{O3}\} \\
&size \in \{1, 2, 3, 4, 5, 6, 7, 8, 9, 10\}
\end{aligned}
\]
For a certain group of $(th,c,size)$ values, the executable program is decided by the above definition. In total, there are $131\times3\times10=3930$ number of choices.

The evaluated objects are an x86 CPU (noted as \texttt{obj1=CPU\_A}) with 14 cores, 32 KiB I/D L1 cache, 1 MiB L2 cache, and 19.25 MiB Unified L3 cache, and another ARM CPU (noted as \texttt{obj2=CPU\_B}) with 32 cores, 64 KiB I/D L1 cache, 512 KiB L2 cache, and 32 MiB Unified L3 cache. Each configuration of the program is run three times (as $r=3$ in DoEs) in case of chance fluctuations such as out of order execution on each CPU, with the output index of the running time. So far, the meta evaluation result of the index and objects is decided and obtained.

Six evaluation methods ($2^k$, $2^kr$, $l^k$, RCTs, do-calculus, and evaluatology methods) compute with the result of every index and every object as the input, and the accuracies and costs of evaluation as the output, as shown in Section~\ref{sec:results}. The detailed computing method is explained in this section right after the explanation of eight tasks.

\subsection{large language model (LLM) problem solving}

As defined in Section~\ref{sec:results}, this task evaluates accuracies and costs of different evaluation methods. The methods evaluates the performance of given LLMs under various configurations of multiple-choice mathematical questions. The questions are 20 multiple choices questions, derived from the MATH-500 dataset~\cite{lightman2023let}, given in natural language and mathematical symbols, most of which have a single correct answer. In this task, LLMs answer many variations of the certain 20 questions, with the changes as follows: the number of shots $sh$, COT prompt $cot$, permutation of choices order $p$, language $lan$, modification of choices $mod$, and role-play prompt $role$.

The EC space is defined as $sh\times  cot\times  p\times lan\times mod\times role$ with given values as follows:\\
\[
\begin{aligned}
&sh \in \{0, 1, 2, 3, 4, 5, 6, 7\} \\
&cot \in \{\text{none}, \text{COT1}\cite{lightman2023let}, \text{COT2}\cite{kojima2022large}, \text{COT3}\cite{zhou2022large}\} \\
&p \in \{\text{none}, \text{swap 2 wrong choices}, \text{swap 1 correct and 1 wrong choices}, \text{swap all}\} \\
&lan \in \{\text{Chinese}, \text{English}, \text{German}\} \\
&mod \in \{\text{none}, \text{random}\} \\
&role \in \{\text{none}, \text{mathematician}, \text{student}, \text{math teacher}\}
\end{aligned}
\]
For a certain group of $(sh,cot,p,lan,mod,role)$ values, the variation of the 20 questions is decided by the above definition. In total, there are $8\times4\times4\times3\times2\times4=3072$ number of variations.

The evaluated objects (LLMs) are \texttt{obj1=gpt-4o-mini}~\cite{bubeck2023sparks}, \texttt{obj2=dschat}~\cite{liu2024deepseek}, and \texttt{obj3=qwen-2.5}~\cite{bai2023qwen}. Each variation of the 20 questions is run three times (as $r=3$ in DoEs) in case of random errors of the LLMs except \texttt{obj2} due to expense control, with the output index of the correctness rate, computed by the number ratio of correctly answered questions and total questions. So far, the meta evaluation result of the index and objects is decided and obtained.

Six evaluation methods ($2^k$, $2^kr$, $l^k$, RCTs, do-calculus, and evaluatology methods) compute with the result of every index and every object as the input, and the accuracies and costs of evaluation as the output, as shown in Section~\ref{sec:results}. The detailed computing method is explained below.

\subsection{Ground-truth and meta evaluation indexes}

Hereby, we establish a structured process for meta-evaluating eight tasks using the concept of evaluatology~\cite{zhan2024evaluatology}. We delineate the evaluation problem, the space of evaluated objects $O$, and the space of evaluation conditions $C$. The fusion of components $C$ and $O$ forms a basic evaluation system that produces various interim variables and outcomes denoted as $m(C,O)$.

Taking Task 5 as an example, different testing algorithms evaluates the randomness of given logistic sequences. According to our notation, $O$ stands for the algorithm space, where $o\in O$ stands for a certain algorithm. $C$ stands for the space of various configurations, i.e., $C=N\times X_0\times r\times|y|$, where $c\in C$ stands for a certain configuration, e.g., $c=(1,0.1,-1.8,10)$. For evaluation of evaluation methods, as is defined, different methods evaluates the ability of multiple randomness test algorithms. In our notation, $m$ stands for a certain method, e.g., $2^k$ DoE, which inputs a set of configurations $c\in C$ and an object $o$, and outputs the evaluation results, with given indexes. In notation, $m(C,O)=i_1,i_2,\cdots$.

This work focuses on two fundamental meta-evaluation indexes: accuracy and cost. We define the ground-truth as average results of all configurations. In notation,\\
\centerline{$\mathbf{GT}_i(o)=\dfrac{\sum_{C}i(c,o)}{\Vert C\Vert}$,}\\
where $\Vert C\Vert$ is the size of the space (as 1440 in Task 5), $i\in I$ is any index in the index space $I$ (as the true rate in Task 5). We define the ground-truth cost of evaluation as the product of $\Vert C\Vert$ and experiment repetition (as $r=3$ in Task 7). Besides the cost, the accuracy is defined as the distance between estimated performance and the ground-truth performance. In notation,\\
\centerline{$\mathbf{cost}(m,C,O)==\Vert C\Vert$,}\\
\centerline{$\mathbf{accuracy}(m,C,O,i)==\mathrm{dist(}\mathbf{GT}_i(o),\mathbf{mean}_{m,i}(o))$,}\\
where $\mathrm{dist}(\cdot)$ is a distance function. In the next several subsections, we provide its concrete computation for every evaluation method.

%Once the accuracy and the cost of a method is defined, it can be computed by experiment. An accuracy-cost curve is defined as the accuracy tendency within the range of cost. The horizontal axis represents the cost and the vertical axis represents the accuracy. When the accuracy is over 90\%, namely the estimated performance of the object covers the ground-truth performance with the 90\% confidence interval, we consider such evaluation feasible. The horizontal line of accuracy=90\% in the result figures and the index C@0.9A in Section~\ref{sec:results} is based on this.

Once the precision and cost effectiveness of a method are determined, they can be quantified through experimentation. An accuracy-cost curve illustrates how accuracy varies across different cost levels. The cost is depicted on the horizontal axis and accuracy on the vertical axis. Evaluation is deemed viable when the accuracy surpasses 90\%, indicating that the estimated performance of the object aligns with the ground-truth performance within a 90\% confidence interval. In the results section, the horizontal line representing accuracy at 90\% and the index C@0.9A are derived from this criterion.

\subsection{Evaluatology method}

We begin by presenting the evaluatology method, which, despite being introduced in this study, stands out as the most straightforward approach in our meta-evaluation framework. According to Wang et al.~\cite{wang2024achieving}, this method is structured into the following four steps.

The cost of evaluation should be firstly decided. The minimum and maximum cost are $r$ and $\Vert C\Vert\times r$ respectively. If no repetition is required, treat as $r=1$. For example, the minimum and maximum cost of Task 5 are $1$ and $1440$, and $\mathbf{cost}=60$ is chosen.

Secondly, make $\mathbf{cost}$ random sampling without replacements, e.g., $60$ times. Compute the average and standard deviation results of all sampled configurations. In notation,\\
\centerline{$\mathbf{mean}_i(o)=\dfrac{\sum_{C_{\text{sample}}}i(c,o)}{\Vert C_{\text{sample}}\Vert},$}\\
\centerline{$\mathbf{std}_i(o)=\sqrt{\dfrac{\sum_{C_{\text{sample}}}\big(i(c,o)-\mathbf{mean}_{i}(o)\big)^2}{|C_{\text{sample}}|}}$,}\\
where $\Vert C\Vert$ is the size of the space (as 1440 in Task 5), $i$ is any index (as the true rate in Task 5).

%Thirdly, make the $\epsilon$ confidence interval estimation using the average and standard deviation. For configurations less than 30, use the Z-score, otherwise use the T-score. If the ground-truth result falls in the interval, note as success, otherwise note as failure.

Thirdly, to conduct $\epsilon$ confidence interval estimation, utilize the average and standard deviation. When dealing with configurations below 30, apply the Z-score; for configurations exceeding 30, opt for the T-score. Identify success if the ground-truth result lies within the interval; otherwise, denote as failure.

%Finally, repeat the second and third steps $K$ times, compute the success rate as the accuracy of the evaluatology method. The result figures with an accuracy-cost curve in Section~\ref{sec:results} is drawn in accordance with this rule. This work takes $\epsilon=95\%$ and $K=50$ in practice.

To determine the accuracy of the evaluatology method, repeat the second and third steps for a total of $K$ iterations. Compute the success rate based on how many times the ground-truth result falls within the confidence interval. The accuracy-cost curve in Section~\ref{sec:results} illustrates this relationship. In AxiaBench, we set $\epsilon$ to 95\% and utilize $K=50$ iterations to ensure robust analysis.

For other methods, we introduce their differences in each step if there is difference. The rest of procedure are the same.

\subsection{Observation method}

%The observation method is mostly same as the evaluatology method. However, there is difference on the second step. After deciding cost, this method does not require a uniform sampling of $\mathbf{cost}$ samples. Instead, we simulate the real-world distribution of all configurations in the EC. In Task 1, 2, and 3, we used power, Gauss, and mixed logarithm distributions respectively according to specific scenarios. This distribution is also used in the SCM method in late parts. The computation of average and standard deviation, and other steps are the same.

The observation method aligns closely with the evaluatology approach, with a key distinction in the second step. Following cost determination, this method does not mandate uniform sampling of $\mathbf{cost}$ samples. Rather, it replicates the real-world distribution of all configurations within the EC. For Task 1, 2, and 3, power, Gaussian, and mixed logarithm distributions are respectively employed based on distinct scenarios. This distribution is also integrated into the SCM method in subsequent stages. The computation of average, standard deviation, and other procedural steps remains consistent across both methods.

\subsection{DoE methods}

We specify the difference of the second step for DoE methods here.

The $2^k$ method takes $2^k$ configurations from the EC space, where $k$ is the dimension of EC space. In each dimension, half values are considered as `low' and half as `high'. It is required to sample one `low' and one `high' value in each dimension, resulting in $2^k$ samples. Taking Task 3 as an example, the EC space is defined above, we provide a possible sampling as follows:\\
\[
\begin{aligned}
&a \in \{10, 50\} \\
&A_{\text{init}} \in \{\text{up centered}, \text{centered}\} \\
&n \in \{5, 100\} \\
&L \in \{5, 6\} \\
&s \in \{1.0, 2.0\}
\end{aligned}
\]
Different from the evaluatology method, the average and standard deviation computation is illustrated by Jain et al.~\cite{jain1990art}.

The $2^{k-p}$ method takes the same sampling space as that of $2^k$. However, it uses of a $\dfrac{1}{2^p}$ part of samples for analysis. For example, suppose $k=5$, then $2^k$ method takes $2^5=32$ configurations, where all `low' and `high' values in the five dimensions are selected. On the contrary, $2^{k-1}$ method takes $2^4=16$ examples, where only one is chosen in the two adjacent configurations (e.g., select one of the following two: (low, low, high, high, \underline{low}), and (low, low, high, high, \underline{high}).

The $2^kr$ method is similar. The only difference is experiment repetition, the average and standard deviation computation is also illustrated by Jain et al.~\cite{jain1990art}. An equivalent form is as follows:\\
\centerline{$\mathbf{mean}_{i}(o)=\mathbb{E}_{C}[\mu(\cdots,k_i,\cdots;j,o)]$ for $f(\cdots,k_i,\cdots;j)>0$,}\\
\centerline{$\mathbf{std}_{i}(o)=\sqrt{\dfrac{\Sigma_{i=1}^{2^k}\Sigma_{j=1}^ry_{ij}^2-2^krq_0^2-\Sigma_{j=1}^{2^k-1}2^kr(\dfrac{1}{2^k}\Sigma_{i=1}^{2^k}S_{ij}\overline{y_i})^2}{2^k(r-1)}}$,}\\
where $k_i$'s are values at different dimensions, $j$ is the repetition, and other symbols are interim variables computed based on the results. This work used a more loose average and standard deviation. In fact, the above results in inferior performance.

As for the $l^k$ DoE method, the difference from $2^k$ is that, separate values in each dimension into $l$ parts instead of two, and take $l^k$ samples for certain. Again we provide a possible sampling of Task 3 as follows ($l=4$):\\
\[
\begin{aligned}
&a \in \{10, 20, 50, 100\} \\
&A_{\text{init}} \in \{\text{up centered}, \text{left centered}, \text{centered}, \text{spread}\} \\
&n \in \{5, 10, 25, 100\} \\
&L \in \{4, 5, 6, 7\} \\
&s \in \{1.0, 1.5, 2.0, 2.5\}
\end{aligned}
\]
The average and standard deviation computation is illustrated by Jain et al.~\cite{douglas2001design} for $l=3$, but others are similar. This work used $l=3$ and $l=4$ only.

\subsection{RCTs method}

%The RCTs method is mostly same as the evaluatology method. However, there is difference on the second step. After deciding cost, take distinct samples for all the objects. The computation of average and standard deviation is the same. Alternately, the RCTs method can also estimate the average discrepancy between two objects by the difference (subtraction). On the third step, if the confidence interval of discrepancy includes the original point (zero point), we assert that the two objects' performances are indistinguishable.

The RCTs method shares similarities with the evaluatology approach, with the only notable difference in the second step. Following cost determination, unique samples are extracted for all objects. The computation of average and standard deviation remains consistent. Additionally, the RCTs method can compute the average discrepancy between two objects through subtraction. In the third step, if the confidence interval of the discrepancy encompasses the original point (zero point), it indicates that the performances of the two objects are statistically indistinguishable.

\subsection{Casual inference methods}

Causal methods include do-calculus and the SCM method. The do-calculus is basically same as the evaluatology method. However, according to the probability formula of do-calculus, the minimum cost is $\Vert C\Vert\times r$, which means high cost and 100\% accuracy. For further explanation, the probability formula is as follows:\\
\centerline{$\sum_w\mathbb{P}(y|t,w)\mathbb{P}(w|\mathbf{do}(t))=\sum_w\mathbb{P}(y|t,w)\mathbb{P}(w)$.}\\
In order to compute the probability and get the distribution of $y$, all the configurations are needed for computation.

The SCM method aligns with the causal model utilized in do-calculus, rendering it equivalent when the entire EC space undergoes evaluation. Nonetheless, distinctions arise when certain configurations are absent. Our study encompasses Task 1 and Task 2, employing identical distribution functions for EC configurations as observed in the observation method. Subsequently, we mask a specific dimension of the EC space and infer the distribution based on SCM theory and the causal diagrams pertinent to the two tasks. Further details are delineated below, with the EC spaces previously defined.

\begin{figure}
    \centering
    \subfigure[The casual diagram of Task 1]{
        \includegraphics[scale=.5]{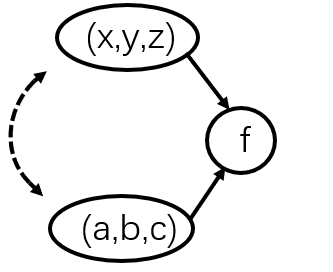}
        \label{fig:task1scm}}
    \qquad
    \subfigure[The casual diagram of Task 2]{
        \includegraphics[scale=.5]{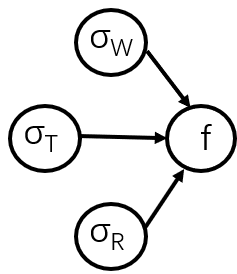}
        \label{fig:task2scm}}
    \caption{Casual diagrams.}
    \label{fig:scm}
\end{figure}

\textbf{Task 1:}\\
We carry on two experiments, by masking $(a,b,c)$ and $(x,y,z)$ respectively. The casual diagram of this task is in Figure~\ref{fig:task1scm}. According to the SCM theory, the distribution of $(a,b,c)$ is delineated by:\\
\centerline{$\mathbb{P}(i|\mathbf{do}((a,b,c)=(a_0,b_0,c_0)))=\sum\limits_{\substack{(x_0,y_0,z_0)\in(x,y,z)}}[\mathbb{P}(i|(a_0,b_0,c_0))\mathbb{P}((a_0,b_0,c_0))]$,}\\
where $(a_0,b_0,c_0)$ and $(x_0,y_0,z_0)$ are values of the corresponding dimension. For the same reason, the distribution of $(x,y,z)$ is delineated by:\\
\centerline{$\mathbb{P}(i|\mathbf{do}((x,y,z)=(x_0,y_0,z_0)))=\sum\limits_{\substack{(a_0,b_0,c_0)\in(a,b,c)}}[\mathbb{P}(i|(x_0,y_0,z_0))\mathbb{P}((x_0,y_0,z_0))]$.}

\textbf{Task 2:}\\
We carry on three experiments, by masking $\sigma_W$, $\sigma_T$ and $\sigma_R$ respectively. The casual diagram of this task is in Figure~\ref{fig:task2scm}. According to the SCM theory, the distribution of $\sigma_W$ is delineated by:\\
\centerline{$\mathbb{P}(i|\mathbf{do}(\sigma_W=\sigma_{W,0}))=\mathbb{P}(i|\sigma_{W,0},\sigma_{T,0},\sigma_{R,0})$,}\\
where $\sigma_{W,0}$, $\sigma_{T,0}$, and $\sigma_{R,0}$ are values of the corresponding dimension. For the same reason, the distribution of $\sigma_T$ and $\sigma_R$ are delineated by:\\
\centerline{$\mathbb{P}(i|\mathbf{do}(\sigma_T=\sigma_{T,0}))=\mathbb{P}(i|\sigma_{W,0},\sigma_{T,0},\sigma_{R,0})$,}\\
and\\
\centerline{$\mathbb{P}(i|\mathbf{do}(\sigma_R=\sigma_{R,0}))=\mathbb{P}(i|\sigma_{W,0},\sigma_{T,0},\sigma_{R,0})$.}

\subsection{Quasi-experiment methods}

The quasi-experiment methods encompass NRCTs and intervention staggering.

NRCTs is applied in Task 4 and Task 6, differing from the RCTs method primarily in the partitioning approach during the second step. While RCTs randomly divides the configuration space into $x$ parts corresponding to the number of objects, NRCTs partitions the space into $x$ parts that are more contiguous and reflective of real-world conditions. The practical partitioning specifics for Task 4 and Task 6 are outlined below, building upon the previously defined EC spaces.

\textbf{Task 4:}\\
\[
\begin{aligned}
&\texttt{obj1}: \text{one half of } M=\text{XYZ} \\
&\texttt{obj2}: \text{another half of } M=\text{XYZ} \\
&\texttt{obj3}: \text{one half of } M=\text{XY}_2\text{Z} \\
&\texttt{obj4}: \text{another half of } M=\text{XY}_2\text{Z}
\end{aligned}
\]

\textbf{Task 6:}\\
\[
\begin{aligned}
&\texttt{obj1}: n \in \{2, 5\},\ \epsilon \in [0.2, 0.4) \\
&\texttt{obj2}: n \in \{10, 100\},\ \epsilon \in [0.2, 0.4) \\
&\texttt{obj3}: n \in \{10, 100\},\ \epsilon \in [0.0, 0.2) \\
&\texttt{obj4}: n \in \{10, 100\},\ \epsilon \in [0.4, 1.0] \\
&\texttt{obj5}: n \in \{2, 5\},\ \epsilon \in [0.4, 1.0] \\
&\texttt{obj6}: n \in \{2, 5\},\ \epsilon \in [0.0, 0.2)
\end{aligned}
\]

Unlike the RCTs and other types of controlled experiments, the intervention staggering method means different intervening time in time sequences to different objects. If such processing does not influence the accuracies of final results, it means this method is feasible for a certain task.

We deploy experiments on Task 2 as the generated precipitation sequences are time sequences and can be manipulated within such requirement. We provide the $t$ ranges for four objects as follows.

\textbf{Task 2:}\\
\[
\begin{aligned}
&\texttt{obj1}: t \in [0, 2190) \\
&\texttt{obj2}: t \in [0, 4380) \\
&\texttt{obj3}: t \in [0, 6570) \\
&\texttt{obj4}: t \in [0, 8760)\ (\text{unchanged})
\end{aligned}
\]

Following the reordering of sequences to align with the respective objects, the algorithms undergo training and testing procedures. The evaluation outcomes are regarded as the outputs of these objects. Notably, the ground-truth average and standard deviations remain consistent with the original results, mirroring the approach adopted in the evaluatology method.

%\subsection{Supplementary information}

%If your article has accompanying supplementary file/s please state so here. 

%Authors reporting data from electrophoretic gels and blots should supply the full unprocessed scans for key as part of their Supplementary information. This may be requested by the editorial team/s if it is missing.

%Please refer to Journal-level guidance for any specific requirements.

\subsection*{Acknowledgements}

We acknowledge supports from the Innovation Funding of ICT, CAS under Grant No. E461070.

\subsection*{Author contribution}

H.L. conceptualized this study, proposed the mathematical foundation and the concrete experiment designs, implemented the experiments of Task 1, 2, 3, 4, 5, and 6, analyzed all the results, and wrote the manuscript. C.W. implemented the experiments of Task 7, collected and analyzed the data, and discussed about the insight. F.F. designed the experiments of Task 8 and analyzed the data, and discussed about the insight. Z.W. collected APIs, implemented the experiments of Task 8 and collected the data. J.Z. put forward the basic problem of this study, provided the basic research methods and revised the manuscript. W.G., and L.W. directed the project, and revised the manuscript. All authors have read and approved the final manuscript.

\bibliographystyle{ieeetr}
\bibliography{sn-bibliography}

@article{zhan2024evaluatology,
  title={Evaluatology: The science and engineering of evaluation},
  author={Zhan, Jianfeng and Wang, Lei and Gao, Wanling and Li, Hongxiao and Wang, Chenxi and Huang, Yunyou and Li, Yatao and Yang, Zhengxin and Kang, Guoxin and Luo, Chunjie and others},
  journal={BenchCouncil Transactions on Benchmarks, Standards and Evaluations},
  volume={4},
  number={1},
  pages={100162},
  year={2024},
  publisher={Elsevier}
}

@book{imbens2015causal,
  title={Causal inference in statistics, social, and biomedical sciences},
  author={Imbens, Guido W and Rubin, Donald B},
  year={2015},
  publisher={Cambridge university press},
  address={Cambridge, UK}
}

@book{thyer2012quasi,
  title={Quasi-experimental research designs},
  author={Thyer, Bruce A},
  year={2012},
  publisher={Oxford University Press},
  address={Oxford, UK}
}

@book{jain1990art,
  title={The art of computer systems performance analysis},
  author={Jain, Raj},
  year={1990},
  publisher={john wiley \& sons},
  address={New York, US}
}

@article{wang2024achieving,
  title={Achieving consistent and comparable CPU evaluation outcomes},
  author={Wang, Chenxi and Wang, Lei and Gao, Wanling and Yang, Yikang and Zhou, Yutong and Zhan, Jianfeng},
  journal={arXiv preprint arXiv:2411.08494},
  year={2024}
}

@article{chalmers1981method,
  title={A method for assessing the quality of a randomized control trial},
  author={Chalmers, Thomas C and Smith Jr, Harry and Blackburn, Bradley and Silverman, Bernard and Schroeder, Biruta and Reitman, Dinah and Ambroz, Alexander},
  journal={Controlled clinical trials},
  volume={2},
  number={1},
  pages={31--49},
  year={1981},
  publisher={Elsevier}
}

@misc{zalta2003stanford,
  title={Stanford encyclopedia of philosophy},
  author={Zalta, Edward N and Nodelman, Uri and Allen, Colin and Perry, John},
  year={2003},
  publisher={Stanford University. The Metaphysics Research Lab Stanford}
}

@book{scriven1991evaluation,
  title={Evaluation thesaurus},
  author={Scriven, Michael},
  year={1991},
  publisher={Sage}
}

@book{lichtenberg2013regular,
  title={Regular and chaotic dynamics},
  author={Lichtenberg, Allan J and Lieberman, Michael A},
  volume={38},
  year={2013},
  publisher={Springer Science \& Business Media},
  address={Berlin, Germany}
}

@article{jumper2021highly,
  title={Highly accurate protein structure prediction with AlphaFold},
  author={Jumper, John and Evans, Richard and Pritzel, Alexander and Green, Tim and Figurnov, Michael and Ronneberger, Olaf and Tunyasuvunakool, Kathryn and Bates, Russ and {\v{Z}}{\'\i}dek, Augustin and Potapenko, Anna and others},
  journal={nature},
  volume={596},
  number={7873},
  pages={583--589},
  year={2021},
  publisher={Nature Publishing Group UK London}
}

@book{cambel1993applied,
  title={Applied chaos theory: A paradigm for complexity},
  author={Cambel, Ali Bulent},
  year={1993},
  publisher={Elsevier},
  address={Amsterdam, Netherlands}
}

@article{burrows1991objective,
  title={Objective guidance for 0--24-hour and 24--48-hour mesoscale forecasts of lake-effect snow using CART},
  author={Burrows, William R},
  journal={Weather and forecasting},
  volume={6},
  number={3},
  pages={357--378},
  year={1991}
}

@article{turing1990chemical,
  title={The chemical basis of morphogenesis},
  author={Turing, Alan Mathison},
  journal={Bulletin of mathematical biology},
  volume={52},
  pages={153--197},
  year={1990},
  publisher={Springer}
}

@book{tilman1997spatial,
  title={Spatial ecology: the role of space in population dynamics and interspecific interactions},
  author={Tilman, David and Kareiva, Peter},
  volume={30},
  year={1997},
  publisher={Princeton University Press},
  address={Princeton, New Jersey, US}
}

@book{wicks1963thermodynamic,
  title={Thermodynamic properties of 65 elements: their oxides, halides, carbides and nitrides},
  author={Wicks, Charles E and Block, Frank E},
  volume={605},
  year={1963},
  publisher={US Government Printing Office},
  address={Washington, D.C., US}
}

@article{kohn1965self,
  title={Self-consistent equations including exchange and correlation effects},
  author={Kohn, Walter and Sham, Lu Jeu},
  journal={Physical review},
  volume={140},
  number={4A},
  pages={A1133},
  year={1965},
  publisher={APS}
}

@article{may1976simple,
  title={Simple mathematical models with very complicated dynamics},
  author={May, Robert M},
  journal={Nature},
  volume={261},
  number={5560},
  pages={459--467},
  year={1976},
  publisher={Nature Publishing Group UK London}
}

@book{rukhin2001statistical,
  title={A statistical test suite for random and pseudorandom number generators for cryptographic applications},
  author={Rukhin, Andrew and Soto, Juan and Nechvatal, James and Smid, Miles and Barker, Elaine and Leigh, Stefan and Levenson, Mark and Vangel, Mark and Banks, David and Heckert, Alan and others},
  volume={22},
  year={2001},
  publisher={US Department of Commerce, Technology Administration, National Institute of~…},
  address={Gaithersburg, Maryland, US}
}

@article{axelrod1981evolution,
  title={The evolution of cooperation},
  author={Axelrod, Robert and Hamilton, William D},
  journal={science},
  volume={211},
  number={4489},
  pages={1390--1396},
  year={1981},
  publisher={American Association for the Advancement of Science}
}

@article{axelrod1988further,
  title={The further evolution of cooperation},
  author={Axelrod, Robert and Dion, Douglas},
  journal={Science},
  volume={242},
  number={4884},
  pages={1385--1390},
  year={1988},
  publisher={American Association for the Advancement of Science}
}

@article{yang2023gpt,
  title={Gpt can solve mathematical problems without a calculator},
  author={Yang, Zhen and Ding, Ming and Lv, Qingsong and Jiang, Zhihuan and He, Zehai and Guo, Yuyi and Bai, Jinfeng and Tang, Jie},
  journal={arXiv preprint arXiv:2309.03241},
  year={2023}
}

@article{liu2024finemath,
  title={Finemath: A fine-grained mathematical evaluation benchmark for chinese large language models},
  author={Liu, Yan and Jin, Renren and Shi, Ling and Yao, Zheng and Xiong, Deyi},
  journal={arXiv preprint arXiv:2403.07747},
  year={2024}
}

@book{douglas2001design,
  title={Design and analysis of experiments},
  author={Douglas, C Montgomery},
  year={2001},
  publisher={John Wiley and Sons Inc}
}

@article{powell2018book,
  title={The Book of Why: The New Science of Cause and Effect. Pearl, Judea, and Dana Mackenzie. 2018. Hachette UK.},
  author={Powell, Stephen},
  journal={Journal of MultiDisciplinary Evaluation},
  volume={14},
  number={31},
  pages={47--54},
  year={2018}
}

@article{ben2000cooperative,
  title={Cooperative self-organization of microorganisms},
  author={Ben-Jacob, Eshel and Cohen, Inon and Levine, Herbert},
  journal={Advances in Physics},
  volume={49},
  number={4},
  pages={395--554},
  year={2000},
  publisher={Taylor \& Francis}
}

@article{pfeiffer2001cooperation,
  title={Cooperation and competition in the evolution of ATP-producing pathways},
  author={Pfeiffer, Thomas and Schuster, Stefan and Bonhoeffer, Sebastian},
  journal={Science},
  volume={292},
  number={5516},
  pages={504--507},
  year={2001},
  publisher={American Association for the Advancement of Science}
}

@article{huisman1999biodiversity,
  title={Biodiversity of plankton by species oscillations and chaos},
  author={Huisman, Jef and Weissing, Franz J},
  journal={Nature},
  volume={402},
  number={6760},
  pages={407--410},
  year={1999},
  publisher={Nature Publishing Group UK London}
}

@book{murray2007mathematical,
  title={Mathematical biology: I. An introduction},
  author={Murray, James D},
  volume={17},
  year={2007},
  publisher={Springer Science \& Business Media},
  address={Berlin, Germany}
}

@article{peterson2021materials,
  title={Materials discovery through machine learning formation energy},
  author={Peterson, Gordon GC and Brgoch, Jakoah},
  journal={Journal of Physics: Energy},
  volume={3},
  number={2},
  pages={022002},
  year={2021},
  publisher={IOP Publishing}
}

@inproceedings{agrawal2016formation,
  title={A formation energy predictor for crystalline materials using ensemble data mining},
  author={Agrawal, Ankit and Meredig, Bryce and Wolverton, Chris and Choudhary, Alok},
  booktitle={2016 IEEE 16th international conference on data mining workshops (ICDMW)},
  pages={1276--1279},
  year={2016},
  organization={IEEE}
}

@article{meredig2014combinatorial,
  title={Combinatorial screening for new materials in unconstrained composition space with machine learning},
  author={Meredig, Bryce and Agrawal, Ankit and Kirklin, Scott and Saal, James E and Doak, Jeff W and Thompson, Alan and Zhang, Kunpeng and Choudhary, Alok and Wolverton, Christopher},
  journal={Physical Review B},
  volume={89},
  number={9},
  pages={094104},
  year={2014},
  publisher={APS}
}

@misc{SPECCPU2017,
author = {{SPEC}},
title  = {{SPEC CPU2017 Benchmark Suite}},
howpublished = "\url{https://www.spec.org/cpu2017/}",
year = {2017}
}

@article{kojima2022large,
  title={Large language models are zero-shot reasoners},
  author={Kojima, Takeshi and Gu, Shixiang Shane and Reid, Machel and Matsuo, Yutaka and Iwasawa, Yusuke},
  journal={Advances in neural information processing systems},
  volume={35},
  pages={22199--22213},
  year={2022}
}

@misc{bubeck2023sparks,
  title={Sparks of artificial general intelligence: Early experiments with gpt-4},
  author={Bubeck, S{\'e}bastien and Chadrasekaran, Varun and Eldan, Ronen and Gehrke, Johannes and Horvitz, Eric and Kamar, Ece and Lee, Peter and Lee, Yin Tat and Li, Yuanzhi and Lundberg, Scott and others},
  year={2023},
  publisher={ArXiv}
}

@article{liu2024deepseek,
  title={Deepseek-v2: A strong, economical, and efficient mixture-of-experts language model},
  author={Liu, Aixin and Feng, Bei and Wang, Bin and Wang, Bingxuan and Liu, Bo and Zhao, Chenggang and Dengr, Chengqi and Ruan, Chong and Dai, Damai and Guo, Daya and others},
  journal={arXiv preprint arXiv:2405.04434},
  year={2024}
}

@article{bai2023qwen,
  title={Qwen technical report},
  author={Bai, Jinze and Bai, Shuai and Chu, Yunfei and Cui, Zeyu and Dang, Kai and Deng, Xiaodong and Fan, Yang and Ge, Wenbin and Han, Yu and Huang, Fei and others},
  journal={arXiv preprint arXiv:2309.16609},
  year={2023}
}

@book{emanuel1994atmospheric,
  title={Atmospheric convection},
  author={Emanuel, Kerry A},
  year={1994},
  publisher={Oxford university press},
  address={Oxford, UK}
}

@incollection{lorenz2017deterministic,
  title={Deterministic Nonperiodic Flow 1},
  author={Lorenz, Edward N},
  booktitle={Universality in Chaos, 2nd edition},
  pages={367--378},
  year={2017},
  publisher={Routledge},
  address={London, UK}
}

@article{rossler1983chaotic,
  title={The chaotic hierarchy},
  author={R{\"o}ssler, Otto E},
  journal={Zeitschrift f{\"u}r Naturforschung A},
  volume={38},
  number={7},
  pages={788--801},
  year={1983},
  publisher={Verlag der Zeitschrift f{\"u}r Naturforschung}
}

@article{dingwell2006lyapunov,
  title={Lyapunov exponents},
  author={Dingwell, Jonathan B},
  journal={Wiley encyclopedia of biomedical engineering},
  year={2006},
  publisher={Wiley Online Library}
}

@article{fasano1987multidimensional,
  title={A multidimensional version of the Kolmogorov--Smirnov test},
  author={Fasano, Giovanni and Franceschini, Alberto},
  journal={Monthly Notices of the Royal Astronomical Society},
  volume={225},
  number={1},
  pages={155--170},
  year={1987},
  publisher={The Royal Astronomical Society}
}

@article{clausius1850ueber,
  title={Ueber die bewegende Kraft der W{\"a}rme und die Gesetze, welche sich daraus f{\"u}r die W{\"a}rmelehre selbst ableiten lassen},
  author={Clausius, Rudolf},
  journal={Annalen der Physik},
  volume={155},
  number={3},
  pages={368--397},
  year={1850},
  publisher={WILEY-VCH Verlag Leipzig}
}

@article{zuur2009glm,
  title={GLM and GAM for count data},
  author={Zuur, Alain F and Ieno, Elena N and Walker, Neil and Saveliev, Anatoly A and Smith, Graham M and Zuur, Alain F and Ieno, Elena N and Walker, Neil J and Saveliev, Anatoly A and Smith, Graham M},
  journal={Mixed effects models and extensions in ecology with R},
  pages={209--243},
  year={2009},
  publisher={Springer}
}

@article{brown2018dieharder,
  title={Dieharder},
  author={Brown, Robert G and Eddelbuettel, Dirk and Bauer, David},
  journal={Duke University Physics Department Durham, NC},
  pages={27708--0305},
  year={2018}
}

@inproceedings{lightman2023let,
  title={Let's verify step by step},
  author={Lightman, Hunter and Kosaraju, Vineet and Burda, Yuri and Edwards, Harrison and Baker, Bowen and Lee, Teddy and Leike, Jan and Schulman, John and Sutskever, Ilya and Cobbe, Karl},
  booktitle={The Twelfth International Conference on Learning Representations},
  year={2023}
}

@inproceedings{zhou2022large,
  title={Large language models are human-level prompt engineers},
  author={Zhou, Yongchao and Muresanu, Andrei Ioan and Han, Ziwen and Paster, Keiran and Pitis, Silviu and Chan, Harris and Ba, Jimmy},
  booktitle={The eleventh international conference on learning representations},
  year={2022}
}

\end{document}